\documentclass[fleqn,usenatbib]{mnras}

\usepackage{newtxtext,newtxmath}

\usepackage[T1]{fontenc}

\DeclareRobustCommand{\VAN}[3]{#2}
\let\VANthebibliography\thebibliography
\def\thebibliography{\DeclareRobustCommand{\VAN}[3]{##3}\VANthebibliography}

\usepackage{graphicx}	
\usepackage{amsmath}	

\title[A massive plume in Abell 2390]{A massive multiphase plume of gas in Abell 2390's brightest cluster galaxy}

\author[Tom Rose et al.]{
Tom Rose$^{1,2}$\thanks{E-mail: thomas.rose@uwaterloo.ca},
B. R. McNamara$^{1,2}$,
F. Combes$^{3}$,
A. C. Edge$^{4}$,
H. Russell$^{5}$,\newauthor
P. Salom\'e$^{3}$,
P. Tamhane$^{1,2}$,
A. C. Fabian$^{6}$,
G. Tremblay$^{7}$,
\\
$^{1}$Department of Physics and Astronomy, University of Waterloo, Waterloo, ON N2L 3G1, Canada \\
$^{2}$Waterloo Centre for Astrophysics, Waterloo, ON N2L 3G1, Canada \\
$^{3}$LERMA, Observatoire de Paris, PSL Research Univ., College de France, CNRS, Sorbonne Univ., Paris, France\\
$^{4}$Centre for Extragalactic Astronomy, Durham University, DH1 3LE, UK\\
$^{5}$School of Physics \& Astronomy, University of Nottingham, Nottingham, NG7 2RD, UK\\
$^{6}$Institute of Astronomy, Cambridge University, Madingley Rd., Cambridge, CB3 0HA, UK\\
$^{7}$Harvard-Smithsonian Center for Astrophysics, 60 Garden St., Cambridge, MA 02138, USA\\
}

\date{Accepted XXX. Received YYY; in original form ZZZ}

\pubyear{2023}

\begin{document}
\label{firstpage}
\pagerange{\pageref{firstpage}--\pageref{lastpage}}
\maketitle

\begin{abstract}
We present new ALMA CO(2-1) observations tracing $2.2 \times 10^{10}\, \textnormal{M}_{\odot}$ of molecular gas in Abell 2390's brightest cluster galaxy, where half the gas is located in a one-sided plume extending 15\,kpc out from the galaxy centre. This molecular gas has a smooth and positive velocity gradient, and is receding 250\,km/s faster at its farthest point than at the galaxy centre. To constrain the plume's origin, we analyse our new observations alongside existing X-ray, optical and radio data. We consider the possibility that the plume is a jet-driven outflow with lifting aided by jet inflated X-ray bubbles, is a trail of gas stripped from the main galaxy by ram pressure, or is formed of more recently cooled and infalling gas. The galaxy's star formation and gas cooling rate suggest the lifespan of its molecular gas may be low compared with the plume's age -- which would favour a recently cooled plume. Molecular gas in close proximity to the active galactic nucleus is also indicated by 250\,km/s wide CO(2-1) absorption against the radio core, as well as previously detected CO(1-0) and HI absorption. This absorption is optically thick and has a line of sight velocity towards the galaxy centre of 200\,km/s. We discuss simple models to explain its origin.
\end{abstract}




\section{Introduction}

Many galaxies throughout the Universe lie in gravitationally bound groups or clusters, with a particularly large galaxy at their centre. These massive ellipticals can have a significant effect on the evolution of other cluster members and the inctracluster medium \citep{McNamara2007, Hlavacek-Larrondo2013, Calzadilla2022}. In particular, the cooling of a cluster's hot X-ray emitting atmosphere can lead to the formation of molecular gas. This may then go on to star formation, or fuel the AGN and ultimately lead to the ejection of powerful radio jets and lobes. These inject huge amounts of energy into the cluster, stifling subsequent cooling of the hot atmosphere and ultimately cutting off star formation in a fuelling and feedback loop \citep{McNamara2012}.

Abell 2390 is a rich, X-ray luminous galaxy cluster at a redshift of $z=0.23$ and with a cooling rate of $200-300\,\textnormal{M}_{\odot}\textnormal{yr}^{-1}$ \citep[][]{Allen2001}. Its brightest cluster galaxy, which we focus on in this paper, is spatially coincident with an array of bright optical emission lines which extend to 25\,kpc on one side \citep{Hutchings2000, Alcorn2023}. On the opposite side of the radio core, optical emission lines are only detected up to around 5\,kpc. 

Radio emission from Abell 2390's brightest cluster galaxy is dominated by the complex and energetic radio source B2151+141. High angular resolution observations show it to have young compact jets and a self-absorbed spectrum in the sub-mm, with half the flux coming from synchrotron emission and half from dust at 850\,$\mu$m \citep{Edge1999, Augusto2006}. On larger scales, evidence of much older radio activity is indicated by X-ray cavities in the cluster's hot atmosphere, as well as $300-600$\,kpc wide radio lobes \citep{Savini2019}. These two epochs of radio activity are orthogonal and likely separated by Gyr timescales, leading to speculation of an intermediate period of radio activity from a precessing continuum source \citep{Alcorn2023}.

During its current epoch of radio activity, Abell 2390's radio continuum source has shown limited variability. \citet{Rose2022} present 15\,GHz data from the Owens Valley Radio Observatory (OVRO) from a timespan of almost a decade. They observe the continuum source with an average interval of 8 days and show a constant flux density dating back to at least Feb 2013. However, although the light curve is flat over this long period, there is peak-to-trough variability of around 20 per cent in excess of the level expected due to noise. Using less frequent observations from the Korean VLBI Network (KVN), \citet[][]{Rose2022} also show that the continuum source is constant at 22\,GHz. Comparison of the OVRO and KVN data with 353\,GHz observations from the SCUBA-2 instrument of the James Clerk Maxwell Telescope (JCMT) show a consistent spectral index of approximately $-0.5$.

In this paper we present new observations of the molecular gas in Abell 2390's brightest cluster galaxy, and discuss them in the context of pre-existing optical, X-ray, and low frequency radio observations. Our new data reveals $2.2 \times 10 ^{10}\, \textnormal{M}_{\odot}$ of molecular gas. Half of this gas is in a heavily one-sided plume which has similar dynamics to the ionised gas seen with optical emission lines. We consider its potential origins, which include an outflowing plume formed as a result of the AGN's historical radio activity, and a plume formed following a gravitational disturbance within the cluster.

Throughout the rest of this paper we use `Abell 2390' in reference the brightest cluster galaxy of the Abell 2390 cluster, unless explicitly mentioned. With observations of stellar absorption lines from the Visible Multi-Object Spectrograph (VIMOS), the redshift of the galaxy is found to be $0.2304\pm0.0001$, giving a recession velocity of $69074\pm30$\,km/s \citep[][]{Hamer2016}. The galaxy is located at RA 21:53:36.5 Dec +17:41:45.0. We assume $\Lambda$ cold dark matter cosmology with $\Omega_{\textnormal{M}} =0.3$ and $\Omega_{\Lambda} = 0.7$, and a Hubble constant of H$_{0}$ = 70 km/s/Mpc.

\section{Observations and data reduction}
\label{sec:observations}
\subsection{ALMA}

\begin{table*}
    \caption{The ALMA and MUSE observations presented in this paper.}
    \centering
    \begin{tabular}{lcccccccccr}
    \hline
    Instrument & Line & FOV & Int. Time & Obs. date & Project & Chan. Width & Beam Size & Cont. Flux & Cont. sens. \\
    & & ($\arcsec$) &  (s) & (yyyy-mm-dd) & & & (") & (mJy) & (mJy/beam)\\
    \hline
    ALMA & CO(1-0) & 59$\arcsec$ & 8014 & 2018-01-07 & 2017.1.00629 & 3.1 km/s & $0.71 \times 0.57$ & 19.8 @ 98.3\,GHz & 0.011 \\
    ALMA & CO(2-1) & 30$\arcsec$ & 4838 & 2022-05-20 & 2021.1.00766 & 1.0 km/s & $0.59 \times 0.56$ & 11.0 @ 192\,GHz & 0.029 \\
    MUSE & H$\alpha$ & 1.0$\arcmin$ & 7200 & 2014-09-22 & 094.A-0115 & 0.15nm & $0.83 \times 0.83$ & -- & -- \\
    \hline
    \end{tabular}
    \label{tab:observations}
\end{table*}

\begin{figure*}
	\includegraphics[width=0.75\textwidth]{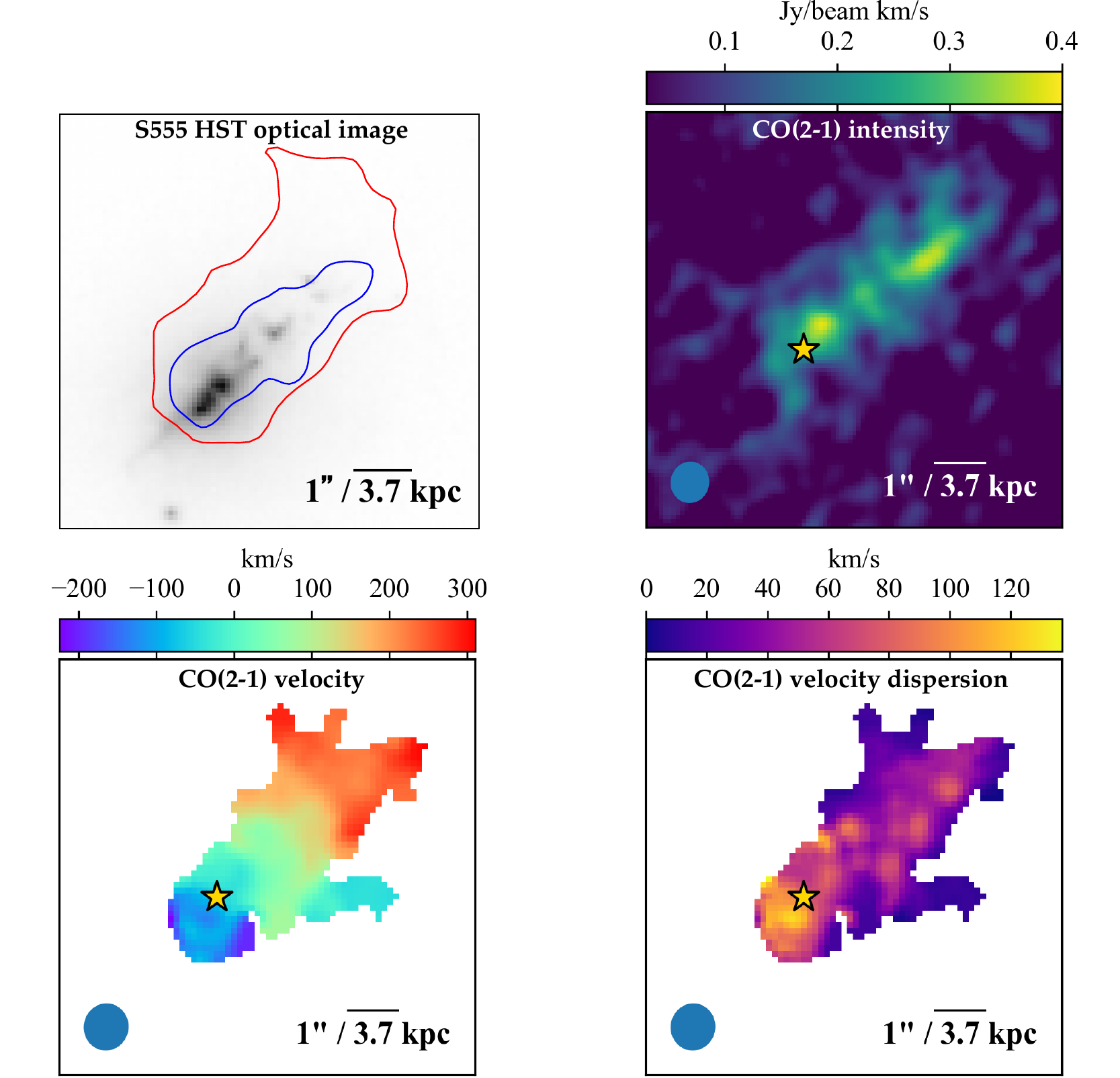}
    \caption{A WFPC2 F555W \textit{HST} image of Abell 2390, alongside moments maps made from its CO(2-1) emission. The blue contour on the optical \textit{HST} image shows the 5\,$\sigma$ outline of the molecular emission. The red contour shows the 5\,$\sigma$ outline of H$\alpha$ emission from MUSE observations, which we discuss later in section \ref{sec:Halpha}. The asymmetry and scale of the molecular emission rule out ordered galaxy rotation as an explanation for the dynamics of the large-scale molecular emission. The velocity is centred on stellar absorption lines of the core \citep{Hamer2016}. The yellow star indicates the location of the radio continuum point source. The CO(2-1) intensity map has a noise level of 0.04\,Jy\,/beam km/s.}
    \label{fig:A2390_HST_and_moments_maps}
\end{figure*}

We present detections of molecular gas in Abell 2390 from two different ALMA observations, details of which are given in Table \ref{tab:observations}. These offer a significant improvement upon the marginal single dish detection of CO(1-0) emission made by \citep{Edge2001}. Our ALMA observation of CO(1-0) emission was taken for project 2017.1.00629.S (PI: A. C. Edge) and CO(2-1) emission was observed for project 2021.1.00766.S (PI: T. Rose). The CO(1-0) observation has previously been shown in \citet[][]{Rose2019b}, where molecular absorption against the galaxy's radio core was discussed, and in \citet[][]{Rose2023}, where intensity, velocity and velocity dispersion maps are shown. We present ALMA's CO(2-1) observations here for the first time, and these offer a significantly improved view of the molecular gas over ALMA's CO(1-0) observations. 

Both ALMA observations were processed with \texttt{CASA} version 6.4.0.16, a software package produced and maintained by the National Radio Astronomy Observatory (NRAO) \citep{CASA}. Self-calibrated measurement sets were provided upon request to The European ALMA Regional Centre, so minimal processing was required to produce our data cubes. From the measurement sets provided, we made continuum subtracted images using the \texttt{CASA} tasks \texttt{tclean} and \texttt{uvcontsub} and a continuum model of polynomial order 1. In \texttt{tclean}, we have used a hogbom deconvolver. We also apply a primary beam correction, use natural weighting and produce images with the narrowest possible channel width.

\subsection{Multi Unit Spectroscopic Explorer}

We present optical observations of Abell 2390 from the Multi Unit Spectroscopic Explorer (MUSE), details of which are given in Table \ref{tab:observations} (project 094.A-0115, P.I.: J. Richard). We use pre-processed data cubes freely available from the ESO Science Archive Facility.

\subsection{\textit{Hubble Space Telescope}}

\textit{Hubble Space Telescope} (\textit{HST}) WFPC2 F555W optical imaging of Abell 2390 was first presented by \citep[][]{Hutchings2000}. In our analysis, we use a pre-processed image freely available from the Hubble Legacy Archive (PropID: 7570).

\subsection{\textit{Chandra} X-ray}

We use \textit{Chandra} X-ray observations of Abell 2390 which have previously been presented in \citet[][]{Allen2001,Sonkamble2015, Shin2016}. The observations date from 1999-11-05, 2000-10-08 and 2003-09-11 and are publicly available from the \textit{Chandra} archive. The three individual observations range between 0.5 and 7.0\,keV, with a combined exposure time of 110\,ks. We repeat the analysis here while considering our new observations of the brightest cluster galaxy's molecular gas. The data were reprocessed using \texttt{ciao} version 4.5 and \texttt{caldb} version 4.6.7. We performed charge transfer inefficiency and time dependent gain corrections on the level 1 event files. Subsequently, we filtered out photons with bad grades. Flare affected periods were identified and removed using the \texttt{lc\_clean} script. We mask bright pixels to remove contamination from point sources in the final image. At distances from the peak flux of between 0.8 and 2.4\,arcmin, we mask pixel values over 5, as well as pixel values above 3 at distances greater than 2.4\,arcmin.

\section{Molecular Gas Properties}

\begin{figure*}
	\includegraphics[width=0.7\textwidth]{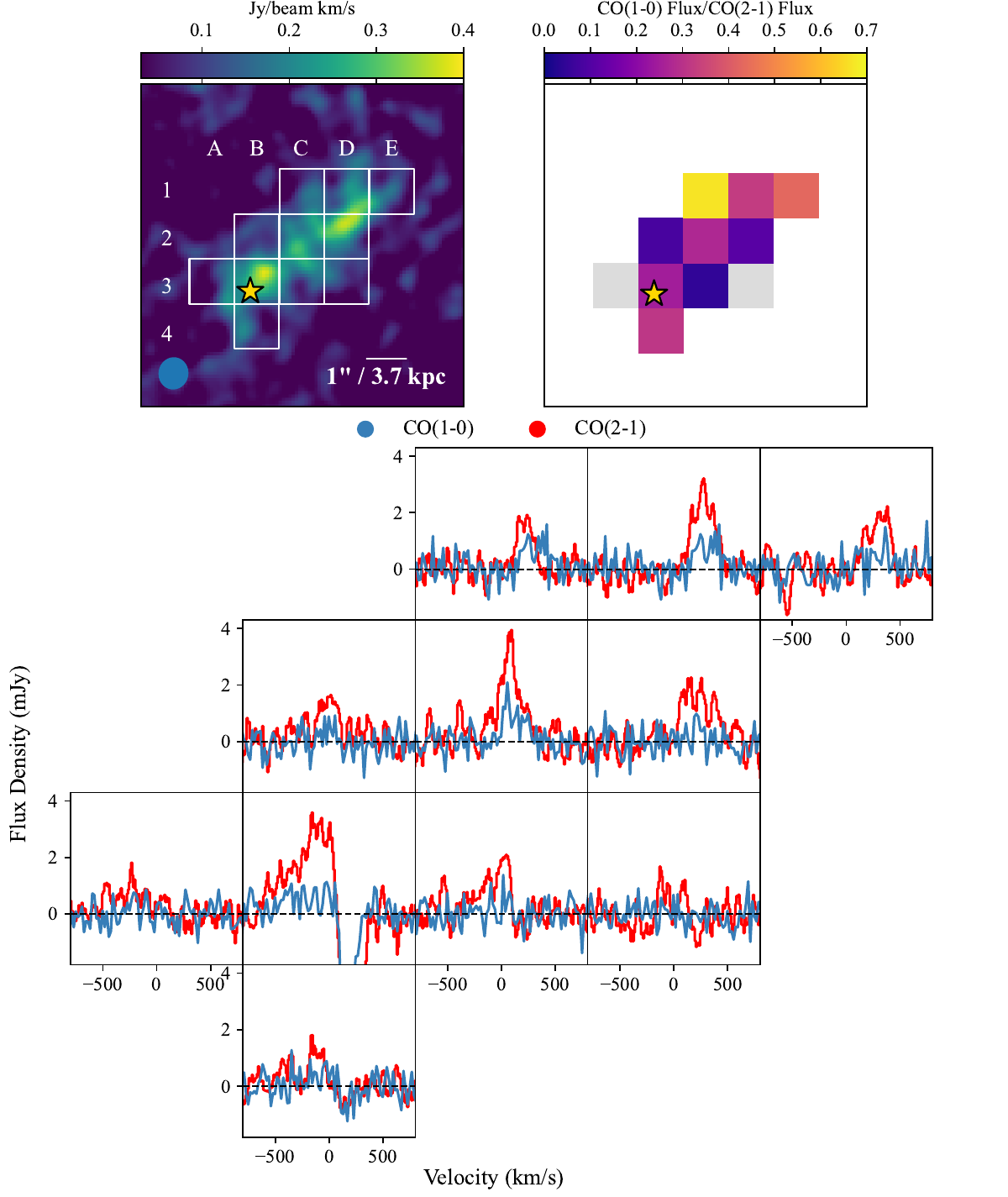}
    \caption{\textbf{Upper:} The CO(2-1) emission in Abell 2390 and the ratio of CO(1-0)/CO(2-1) emission. \textbf{Lower:} CO(1-0) and CO(2-1) spectra of Abell 2390’s molecular emission. Lower CO(1-0)/CO(2-1) ratios imply colder, denser gas. Grey tiles indicate insufficient CO(1-0) emission. Spectra extracted from the line-of-sight to the galaxy’s radio core show molecular absorption. The CO(2-1) absorption is shown in more detail in Fig. \ref{fig:CO(2-1) spectra} and is discussed further in section \ref{sec:absorption}. The yellow star indicates the location of the central continuum source.}
    \label{fig:A2390_emission_grid}
\end{figure*}

In Fig. \ref{fig:A2390_HST_and_moments_maps}, we show intensity, velocity and velocity dispersion maps (moments 0, 1 and 2) produced from ALMA's CO(2-1) observations of Abell 2390. We also reproduce the \textit{Hubble Space Telescope} (\textit{HST}) image first presented by \citet[][]{Hutchings2000} to show how the spatial distribution of molecular and optical emission compare. We do not show intensity, velocity and velocity dispersion maps made with the CO(1-0) data because the line is significantly weaker than the CO(2-1), but they can be seen in \citet[][]{Rose2023}.

Abell 2390's molecular emission appears to have two main components. The brightest is spheroidal and centred slightly to the north-west of the galaxy's radio core. This component does not appear to be spatially resolved in either CO(1-0) or CO(2-1). To the south east of the central component, minimal CO(2-1) is visible. Weak CO(1-0) may be present, but we are unable to confidently state this as a true detection, rather than simply being noise. However, on the same side as the bright spheroidal component of emission is a 15 kpc long lane of molecular gas. This is spatially coincident with the emission seen in the F555W image from \textit{HST} \citep[][]{Edge1999} and has a smooth velocity change from close to zero at the radio core, to around 250 km/s at the farthest point. It is also spatially coincident with an H$\alpha$ plume shown by \citet[][]{Hutchings2000}, \citet{Hamer2016} and \citet{Alcorn2023}, which we also discuss later in section \ref{sec:Halpha}. 

In Fig. \ref{fig:A2390_emission_grid} we also show spatial variation in the ratio of CO(1-0) and CO(2-1) integrated flux densities. The CO(1-0)/CO(2-1) ratio is heavily dependent on the density of the gas, with lower ratios implying colder, denser gas \citep{Penaloza2017}. This has previously been done for many massive galaxies \citep[e.g.][]{Edge2001, SalomeCombes2003}, but ALMA now allows us to achieve some degree of spatial resolution in Abell 2390. There is some indication that this ratio increases towards the farthest point of the plume. This implies less collisional excitation of the molecular gas and therefore a lower density. Radiation from active galactic nuclei may also affect the excitation conditions in some instances \citep{Ivison2012}.

\begin{figure*}
	\includegraphics[width=0.9\textwidth]{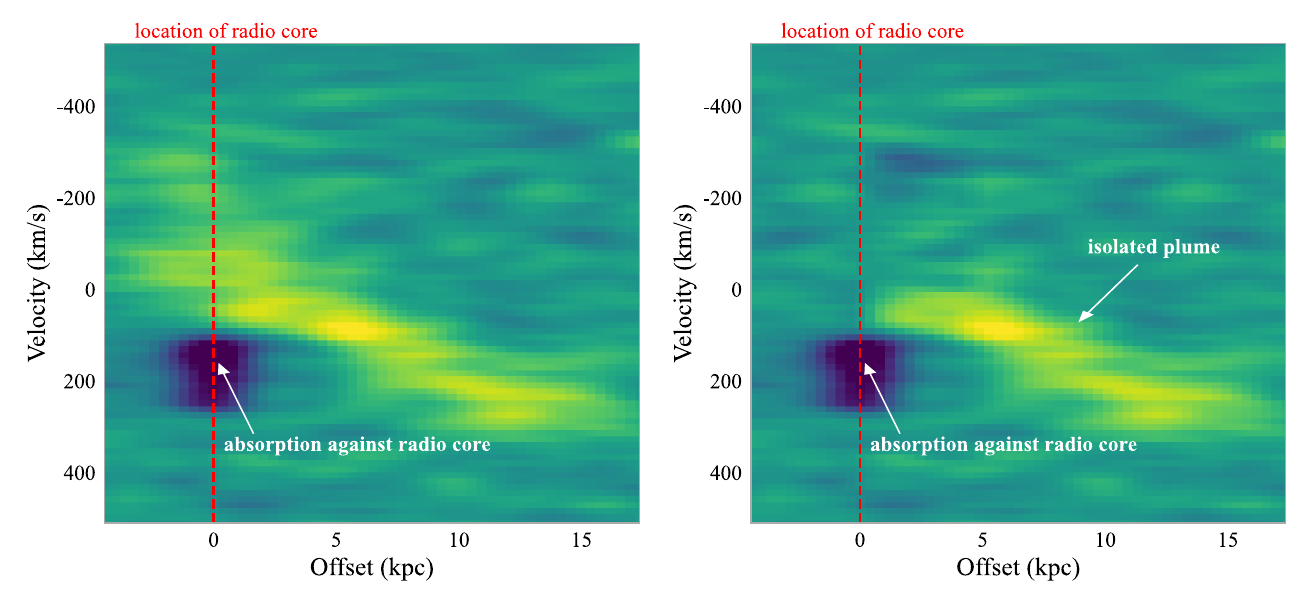}
    \caption{Position-velocity diagram made along the axis of the molecular plume, with a 2.5" width and the zero spatial offset set as the location of the radio continuum source. There appears to be two main components. One is tightly bound to the galaxy and centred close to the continuum source. The other is the 15\,kpc long molecular plume. On the right hand side we show the position-velocity of the isolated molecular plume. To make this we assume emission at negative offsets is due to gas tightly bound to the galaxy and can be treated as a separate component. We then subtract a reflection of this emission about the 0 kpc offset. We restrict the velocity range of this plot, but as can be seen in Fig. \ref{fig:CO(2-1) spectra}, some faint emission from the core is present at up to -700\,km/s. The zero velocity is set by stellar absorption lines of the core \citep{Hamer2016}.}
    \label{fig:pV_diagram}
\end{figure*}

\subsection{Velocity structure of the molecular gas}

Fig. \ref{fig:A2390_HST_and_moments_maps} reveals very clear velocity structure in Abell 2390's molecular gas. Aside from the dense gas close to the radio core, to the north-west (top-right) there is the plume with a velocity increasing linearly from approximately 0 to 250 km/s over a distance of 15\,kpc. All of the velocity structure is along the major axis of this plume, with very little change seen across its 5\,kpc width. As we discuss later, this velocity structure is similar to that seen in optical emission lines, though is less extended \citep{Hutchings2000, Alcorn2023}.

In Fig. \ref{fig:pV_diagram} we show a position-velocity diagram made along the major axis of the plume. On the right, we also show a position-velocity diagram in which we isolate the plume of molecular gas from what appears to be a separate component centred close to the continuum source. To do this, we assume the central component is symmetric, and subtract a reflection of the emission in the position-velocity diagram along the axis shown by the dashed red line. To the left of the line, we replace the emission with noise simulated based on an emission free region. The position-velocity diagram shown in Fig. \ref{fig:pV_diagram} is unusual, both in its asymmetry and the highly uniform change in velocity at increasing distances from the radio core. This steady velocity increase out to radii of at least 15\,kpc indicates a force acting on the gas far from the galaxy centre. Typically, position-velocity diagrams flatten out at larger radii. In the case of a galaxy disc, this is due to the decreased orbital velocities at larger radii. For outflows, it is because the jets lose their power beyond distances of a few kpc.

Fig. \ref{fig:A2390_HST_and_moments_maps} shows that across most of the plume, the velocity dispersion falls between 30 and $60$\,km/s, with several unresolved peaks. The region around the radio core has highly asymmetric dispersion. On the side of the extended plume the dispersion is around 60 km/s, but the value is twice as high on the opposite side.

\subsection{Molecular Mass Estimates}

The molecular mass associated with CO emission can be estimated using the following relation from \citet{Bolatto2013},

\begin{equation}
\begin{split}
\textnormal{M}_{\text{mol}} = \frac{1.05\times 10^{4}}{F_{ul}} \left( \frac{X_{\text{CO}}}{2\times 10^{20}\frac{\text{cm}^{-2}}{\text{K km s}^{-1}}}\right)\left( \frac{1}{1+z}\right) \\ \times \left( \frac{S_{\text{CO}} \Delta v}{\text{Jy km s}^{-1}}\right) \left( \frac{D_{\text{L}}}{\text{Mpc}}\right)^{2}\, \textnormal{M}_{\odot},
\end{split}
\label{eq:massequation}
\end{equation}
where $M_{\text{mol}}$ is the mass of molecular hydrogen, $X_{\text{CO}}$ is a CO-to-H$_{2}$ conversion factor, $z$ is the source's redshift, $S_{\text{CO}} \Delta v$ is the emission integral and $D_{\text{L}}$ is the luminosity distance in Mpc. $F_{ul}$ is an approximate conversion factor for the expected flux density ratios of the CO(1-0) and CO(2-1) lines, where $u$ and $l$ represent the upper and lower levels. For CO(1-0), $F_{10}=1$ and for CO(2-1), $F_{21}=3.2$ originates from a combination of the factor of two between the frequencies of the lines and the brightness temperature ratio observed for molecular clouds in spiral galaxies of 0.8 at 10\,K \citep{BraineandCombes1992}. It is also consistent with the regions with the strongest emission in Fig. \ref{fig:A2390_emission_grid} and is the same as that used by similar studies \citep[e.g.][]{David2014,Tremblay2018}.

The CO-to-H$_{2}$ conversion factor is a considerable source of uncertainty, but to ensure our mass estimates are comparable with similar studies, we use the standard Milky Way value of \mbox{$X_{\text{CO}} = 2 \times 10^{20}$ cm$^{-2}$ (K km s$^{-1}$)$^{-1}$} in our calculations \citep{Bolatto2013}.

From the CO(2-1) emission, we estimate the total molecular mass to be $(2.2\pm0.2)\times 10 ^{10}\, \textnormal{M}_{\odot}$. In Table \ref{tab:masses}, we also show the molecular masses for each of the grid regions in Fig. \ref{fig:A2390_emission_grid}. The core region contains the most concentrated molecular gas and makes up just under a third of the total, with nearly all the rest being in the more extended plume. The total mass we find is consistent with the value from the CO(1-0) emission of $(2.2\pm0.6)\times 10 ^{10}\, \textnormal{M}_{\odot}$ \citep{Rose2023}.

\begin{table}
	\centering
	\caption{The molecular mass contained in the grid regions shown in Fig. \ref{fig:A2390_emission_grid}, determined from the CO(2-1) emission}
	\label{tab:masses}
	\begin{tabular}{lccc} 
		\hline
		Region & v$_{\textnormal{cen}}$ (km/s) & FWHM (km/s) & M$_{\textnormal{mol}}$ (M$_{\odot}$) \\
		\hline
		Total & & & (2.2$\pm$0.2)$\times 10^{10}$ \\
		A3 & -260$\pm$30 & 430$\pm$50 & (1.4$\pm$0.2)$\times 10^{9}$ \\
		B2 & -10$\pm$30 & 240$\pm$40 & (1.3$\pm$0.3)$\times 10^{9}$ \\
		B3 & -90$\pm$40 & 580$\pm$110 & (6.3$\pm$1.2)$\times 10^{9}$ \\
		B4 & -120$\pm$40 & 160$\pm$40 & (0.9$\pm$0.2)$\times 10^{9}$ \\
		C1 & 220$\pm$20 & 160$\pm$20 & (1.2$\pm$0.1)$\times 10^{9}$ \\
		C2 & 70$\pm$20 & 230$\pm$20 & (2.7$\pm$0.3)$\times 10^{9}$ \\
		C3 & -40$\pm$30 & 250$\pm$30 & (1.6$\pm$0.2)$\times 10^{9}$ \\
		D1 & 280$\pm$20 & 230$\pm$20 & (2.4$\pm$0.2)$\times 10^{9}$ \\
		D2 & 220$\pm$20 & 350$\pm$20 & (2.3$\pm$0.2)$\times 10^{9}$\\
		D3 & -100$\pm$40 & 90$\pm$30 & (0.4$\pm$0.2)$\times 10^{9}$ \\
		E1 & 300$\pm$20 & 220$\pm$20 & (1.7$\pm$0.2)$\times 10^{9}$ \\
		\hline
	\end{tabular}
\end{table}

\section{Multiwavelength Observations of Abell 2390}
\subsection{MUSE}
\label{sec:Halpha}

\begin{figure*}
	\includegraphics[width=0.9\textwidth]{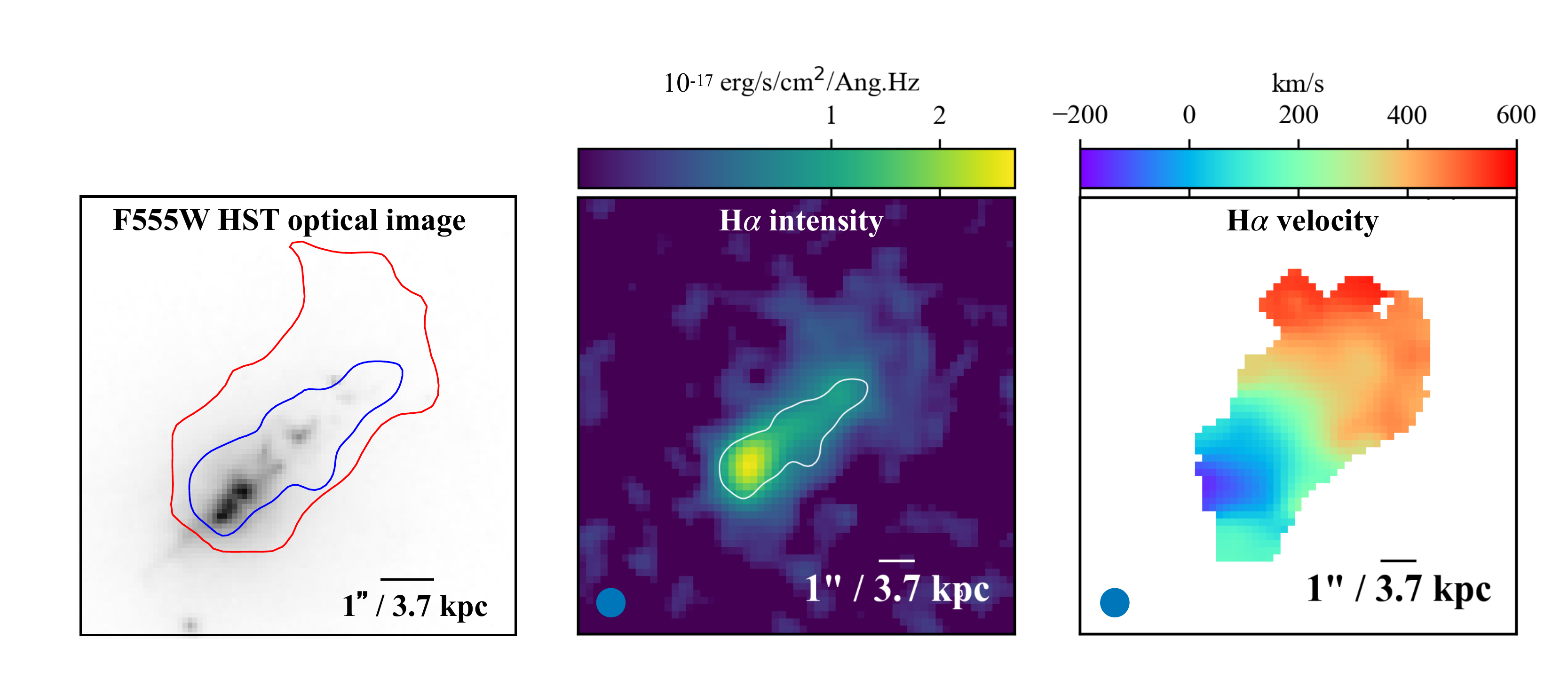}
    \caption{F555W \textit{HST} image alongside H$\alpha$ intensity and velocity maps from the MUSE instrument of the VLT. The blue contour on the optical \textit{HST} image and white contour on the H$\alpha$ image show the 5\,$\sigma$ outline of the molecular emission. The red contour shows the 5\,$\sigma$ outline of H$\alpha$ emission.}
    \label{fig:A2390_HST_and_MUSE_maps}
\end{figure*}

Recently, \citet{Alcorn2023} presented high angular resolution maps of optical emission lines in Abell 2390 from the CFHT/SITELLE imaging Fourier Transform Spectrograph. In Fig. \ref{fig:A2390_HST_and_MUSE_maps} we show moments maps made with similar data from MUSE. These are made using the strongest emission line, H$\alpha$. This emission seems to trace similar structures to the molecular gas, albeit on a slightly larger scale -- with the far extent of the H$\alpha$ plume extending to 25\,kpc and receding 600\,km/s faster than the galaxy centre. Interestingly, to the north of the most redshifted edges of the plume is a `hook' in the H$\alpha$ emission, where the emission broadens and is more redshifted \citep[this is also seen in N{\small II},][]{Alcorn2023}.

\begin{figure}
\centering
	\includegraphics[width=0.9\columnwidth]{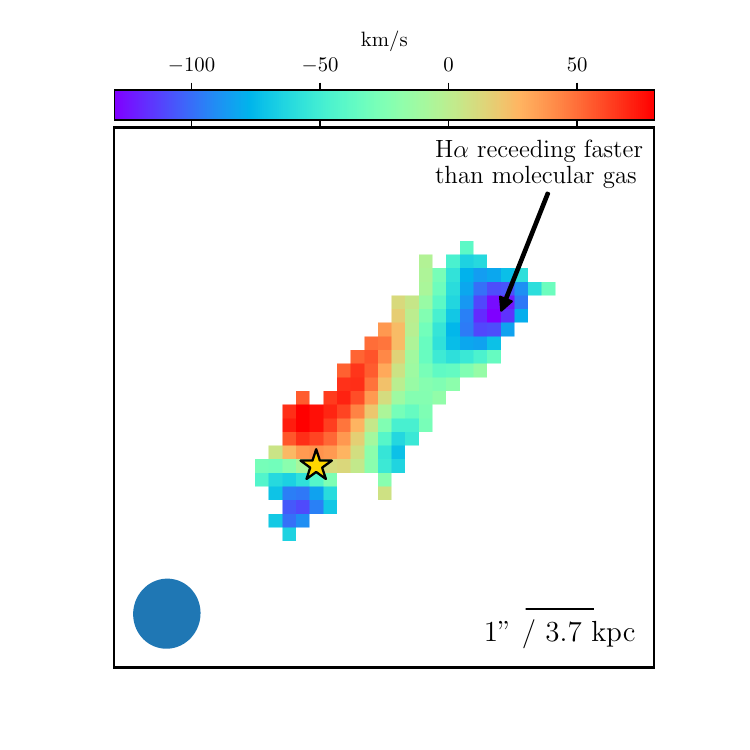}
    \caption{Molecular gas velocity minus H$\alpha$ velocity. In a central spine along the plume the two gas phases have similar velocities. However, to the north of the continuum source, the molecular gas is receding faster. To the south of the continuum source and at the far north-western extent of the molecular plume, the H$\alpha$ emitting gas is receding faster. The yellow star indicates the location of the central continuum source.}
    \label{fig:Molecular_minus_halpha_velocity}
\end{figure}

To compare the two, in Fig. \ref{fig:Molecular_minus_halpha_velocity} we show the molecular gas velocity minus the H$\alpha$ velocity. Along a central spine in the plume, the two gas phases have very similar velocities. However, there are velocity differences to the north and south of the continuum source, and at the far extent of the molecular plume. For the most part, these velocity differences are between a few tens and 100\,km/s. This is around a third to half the velocity dispersion of the H$\alpha$ gas \citep[][fig. 10]{Alcorn2023}, so although there is some discrepancy, it does not imply uncoupled motion. The small differences present may also be due to dust obscuration in the H$\alpha$ emission.

\subsection{\textit{Chandra} X-ray}
\label{sec:Xray}

\begin{figure*}
	\includegraphics[width=0.7\textwidth]{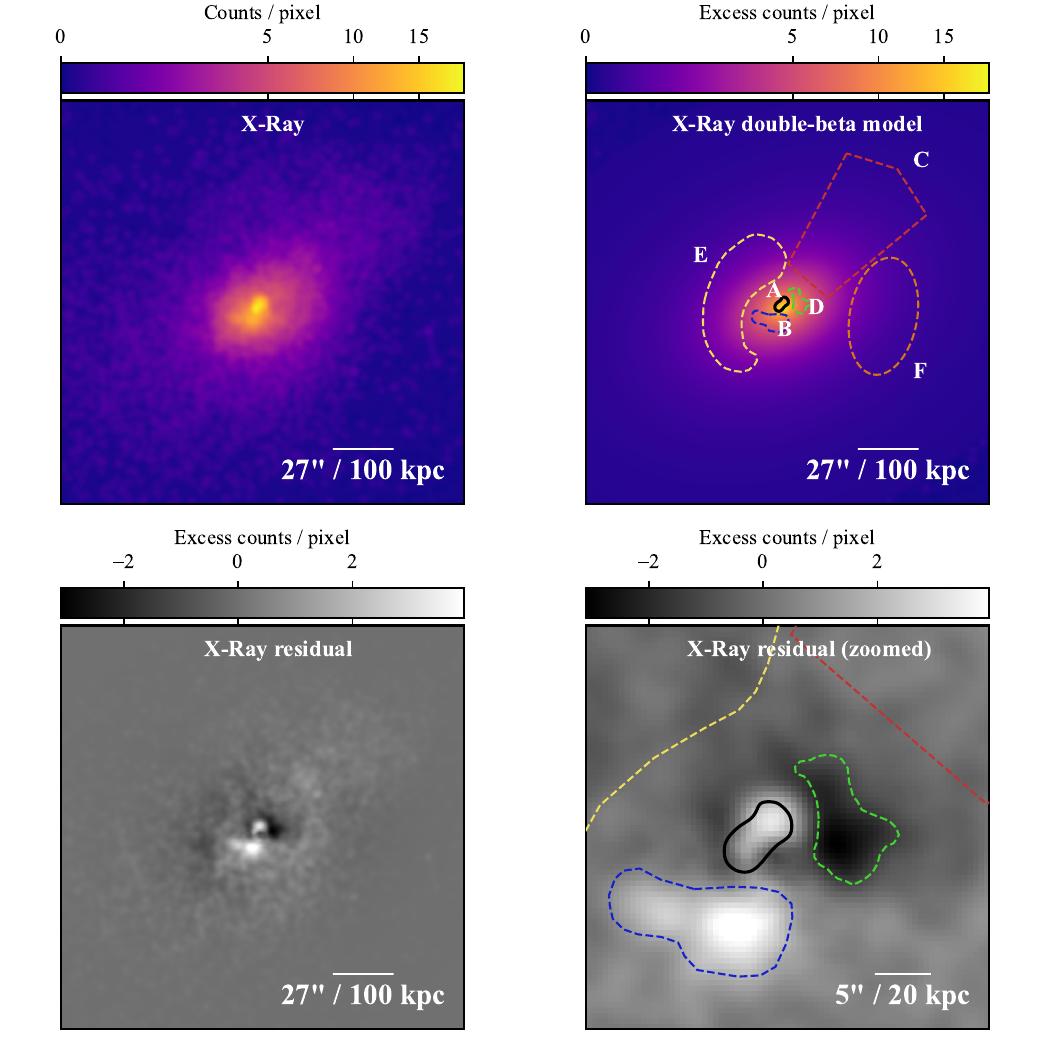}
    \caption{Combined \textit{Chandra} image and the residual X-ray emission after subtraction of a double-beta model fit, all with 2" smoothing. The best-fit location for this model is 7\,kpc from the radio continuum source. The black contour labelled `A' shows the location of the H$\alpha$ emission, while other regions trace the excesses and depressions. Within this region, the residual X-ray emission implies an excess hot gas mass of $2\times 10^{10}\, \textnormal{M}_{\odot}$.}
    \label{fig:X-ray}
\end{figure*}

\begin{figure}
	\includegraphics[width=\columnwidth]{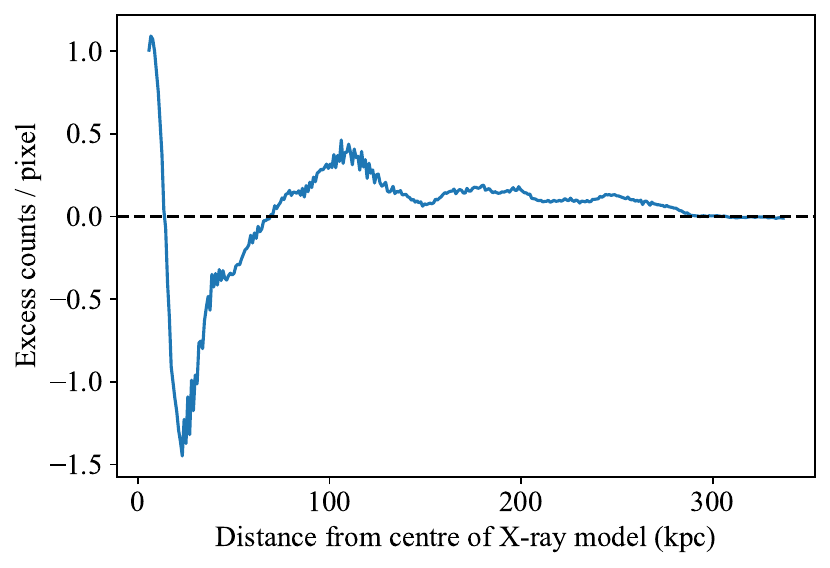}
    \caption{The mean excess counts per pixel as a function of distance from the centre of the X-ray model, for the red region labelled C shown in Fig. \ref{fig:X-ray}. An excess count is calculated for each individual pixel, but here we show the data in annular bins 1 pixel wide. Between around 10 and 70 kpc, there is a cavity, followed by an excess of emission out to around 300 kpc. The implied excess of hot gas in region `C' is $5.9\times 10^{9}\, \textnormal{M}_{\odot}$.}
    \label{fig:X-ray_plume}
\end{figure}

In Fig. \ref{fig:X-ray}, we show an X-ray image of Abell 2390, and a residual after fitting a double-beta model, described by \citet{Ettori2000}. This model represents one cool and one ambient phase, where $S$ is the surface brightness, $R$ is the projected radius on the plane of the sky, r is the core radius and $\beta$ is the density slope parameter. When fitting to the model, we apply a mask to radii less than six pixels (11kpc) from the centre of the fit. We find that the peak brightness of the X-ray emission also gives the best fit double-beta model, i.e. the one which minimizes the sum of the squares of the residuals. This point is approximately 7 kpc from the location of the radio continuum source. The best fit parameters we find are: $S_{\textnormal{cool}}(0) = 6.4$\,counts/pixel, $r_{\textnormal{cool}} = 71$\,kpc, $\beta_{\textnormal{cool}} = 1.2$, $S_{\textnormal{amb}}(0) = 12.6$\,counts/pixel, $r_{\textnormal{amb}} = 7.3$\,kpc and $\beta_{\textnormal{amb}} = 0.46$.

Following subtraction of the double-beta model, the residual image (Fig. \ref{fig:X-ray}) contains three points of excess and and three points of depression \citep[previous analysis by][using a smoothed image as the X-ray model identified five X-ray cavities]{Sonkamble2015}. 

One region of X-ray excess (labelled A) is coincident with the molecular and H$\alpha$ plume and implies an excess mass of $2\times 10^{10}\, \textnormal{M}_{\odot}$, based on a cluster mass of $6.4\times 10^{14}\, \textnormal{M}_{\odot}$ \citep{Allen2001}. The second (labelled B), to the south west of the galaxy centre implies an excess mass of $1.6\times 10^{11}\, \textnormal{M}_{\odot}$. A third (labelled C) exists in a large, roughly 300 kpc wide region and has an implied excess mass of $5.9\times 10^{9}\, \textnormal{M}_{\odot}$. This last region of excess seems to have a similar orientation to the molecular and H$\alpha$ plumes. In Fig. \ref{fig:X-ray_plume} we show the mean excess counts/pixel for this region as a function of radius from the centre of the double-beta model. This shows a large excess in the X-ray surface brightness close to the core, followed by a depression up to around 70\,kpc. Between around 70 and 300\,kpc, there is an excess of X-ray emission. This excess peaks at around 100\,kpc, and seems to oscillate with three peaks and two depressions. This could represent the characteristic spiral patterns in the X-ray emission of `sloshing' galaxy clusters predicted by simulations \citep[e.g.][fig. 5]{ZuHone2010}. The central excess B and depression D may also be a manifestation of this effect.

The strongest and most compact depression (labelled D) lies just beyond the edge of the molecular and H$\alpha$ plume (the outline of the H$\alpha$ plume is indicated in Fig. \ref{fig:X-ray} and labelled A). A second, more extended depression (labelled E) lies to the left of the galaxy centre. To the right of the galaxy centre there is a third depression (labelled F). This is weak in magnitude but extends across around 300\,kpc.

To estimate the significance of the depressions and excesses we use an approach based on that of \citet{Martz2020}. For each depression or excess, the residual image's integrated counts, $N_{I}$, is compared with the model image's integrated counts, $N_{M}$. 

The signal, $S$, in a region is:
\begin{equation}
    S = |N_{I} - N_{M}|
\end{equation}

For depressions, the deficit, $D$, between the image and model is:
\begin{equation}
    D = 1 - \frac{N_{I}}{N_{M}}
\end{equation}

For regions of excess emission, the excess, $E$, between the image and model is:
\begin{equation}
    E = \frac{N_{I}}{N_{M}} - 1
\end{equation}

The signal-to-noise ratio for the cavities and regions of excess is:
\begin{equation}
    \textnormal{SNR} = \frac{S}{\sqrt{S + 2N_{M}}}
\end{equation}

Parameters calculated using the equations above are given in Table \ref{tab:X-ray_significances}. The X-ray excess coincident with the plume is relatively weak at 12 per cent, and has a signal-to-noise ratio of 2.6. The large cavity labelled `F' in Fig. \ref{fig:X-ray} is similarly weak, with a deficit of 13 per cent. Its signal-to-noise ratio is more significant at 5.0. The remaining cavities and excesses we identify are stronger (between 23 and 35 per cent) and have much higher signal-to-noise ratios (at least 7). There is therefore a possibility the excess coincident with the plume is spurious, but all other excesses and deficits are of high significance. 

We note that our X-ray analysis is carried out on a full band (0.5 - 7\,keV) image due to the limited counts. The hard X-ray is core dominated, but the soft X-ray image has some extended emission to the north-west of the continuum. Therefore, apparent deficits may arise due to our radial averaging analysis, resulting in a strong excess and a clear deficit on opposite sides of the centre. In our data the extent of the soft X-ray excess is only around 3" (11\,kpc), so we do not believe this effect to be responsible for the quasi-symmetric excess B and deficit D.

\begin{table}
    \centering
    \begin{tabular}{cccc}
    \hline
        Region & Label & Excess/Deficit (\%) & SNR \\
        \hline
        Plume & A & 12 & 2.6 \\
        Central Excess & B & 35 & 12 \\
        Large Excess & C & 38 & 17 \\
        \hline
        Central depression & D & 26 & 7.0 \\
        Left depression & E & 23 & 14 \\
        Right depression & F & 13 & 5.0 \\
        \hline
    \end{tabular}
    \caption{The magnitude of each region of excess and depression shown in Fig. \ref{fig:X-ray}, plus their signal-to-noise ratios.}
    \label{tab:X-ray_significances}
\end{table}

\section{Plume orientation, power and age}
\subsection{Plume Orientation}
\label{sec:plume_orientations}

\begin{figure*}
	\includegraphics[width=0.9\textwidth]{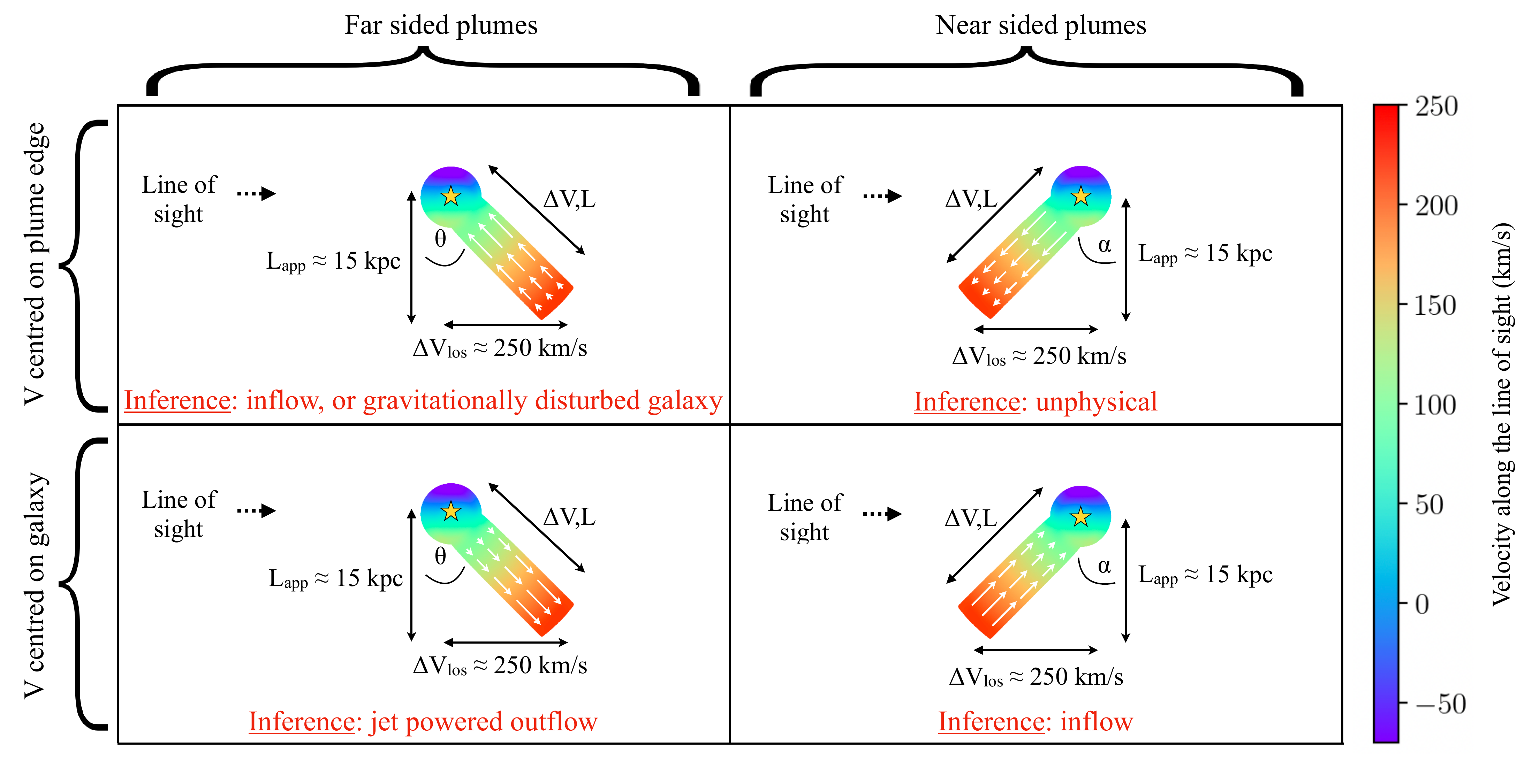}
    \caption{Schematic diagrams showing a bird's eye view of potential plume orientations. We include two basic components: a rotating spherical distribution of gas centred on the radio core, and the 15\,kpc long molecular plume. We consider arrangements where the plume is on the near or far side of the main galaxy, but only the latter are feasible. Positive velocities imply movement away from us along the line of sight relative to the main galaxy. Negative velocities imply movement towards us along the line of sight. Arrows indicate the direction and magnitude of flow.}
    \label{fig:A2390_schematic}
\end{figure*}

\begin{figure}
	\includegraphics[width=0.9\columnwidth]{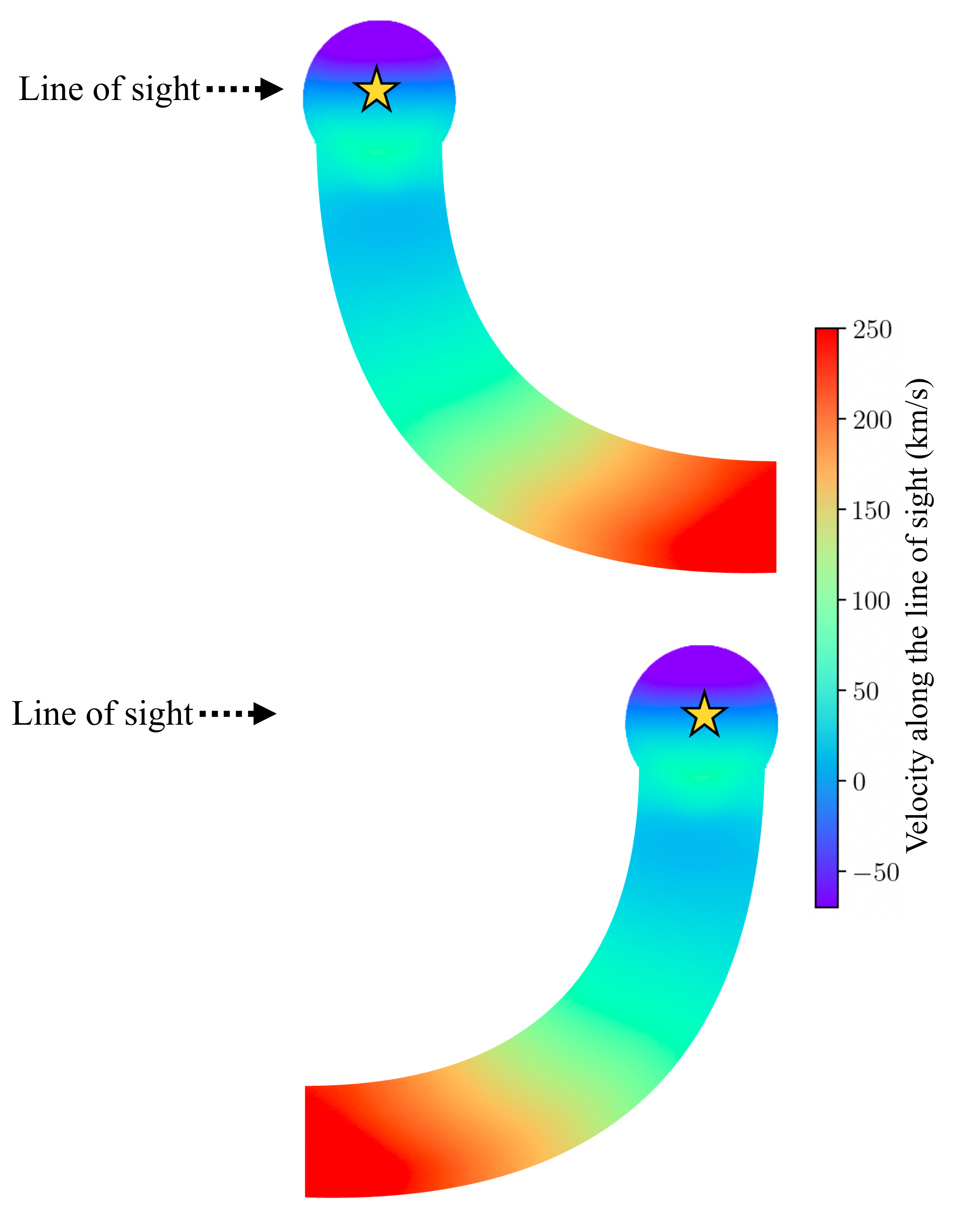}
    \caption{A curved plume of molecular gas with a constant velocity magnitude may theoretically have a smoothly changing line of sight velocity. The far sided plume in the upper diagram shows how a line of sight velocity change may arise from curvature in a jet powered outflow, or an outflow due to a gravitational disturbance. The lower diagram shows how curvature in a near sided plume may cause an apparent decrease in the velocity of inflowing gas.}
    \label{fig:A2390_schematic_curved_plume}
\end{figure}

Since we are seeing the plume in projection, it is likely to lie on either galaxy's far side or near side. Each of the two diagrams on the left of Fig. \ref{fig:A2390_schematic} shows bird's-eye-view of a far sided, linear molecular plume. If the farthest point of the molecular plume is treated as the rest-frame, it is implied that the central galaxy has been accelerated towards us along our line of sight and has left behind a trail of molecular gas. For the lower left arrangement of Fig. \ref{fig:A2390_schematic}, the main galaxy is treated as the rest-frame. The plume can then be interpreted as an outflow induced by an accelerating jet. 

The two arrangements on the right of Fig. \ref{fig:A2390_schematic} show a near sided, linear molecular plume. When the farthest point of the molecular plume is treated as the rest-frame (upper right of diagram) this again implies the central galaxy has been accelerated away from the initial velocity frame and left behind a trail of molecular gas. However, this would require the galaxy to have been pulled from the far side, meaning it should be receding faster than the plume. Another interpretation of this arrangement which may seem feasible on first inspection is a near sided outflow. However, this is again unphysical because the plume's farthest point is in fact receding faster than the main galaxy.

The lower-right arrangement implies the inflow of a near-sided plume. However, the velocity structure of the molecular gas requires that in this case there is deceleration of the gas which is moving towards the main galaxy. This may be possible if the plume is significantly curved, which we show in Fig. \ref{fig:A2390_schematic_curved_plume}.

The velocity structure we see in the molecular plume therefore indicates that Abell 2390's plume lies on the galaxy's far side. This is consistent with the \textit{HST} data in Fig. \ref{fig:A2390_HST_and_moments_maps}, which shows no obvious dust obscuration.

It is also interesting to consider the angle the plume makes to our line of sight. In Fig. \ref{fig:A2390_schematic}, L represents the true length of the plume and L$_{\textnormal{app}}$ is the apparent length, while $\Delta$V represents the total velocity change along the plume and $\Delta$V$_{\textnormal{los}}$ is the velocity change seen along the line of sight. With $\theta$ as defined in Fig. \ref{fig:A2390_schematic}, then 

\begin{equation}
    \Delta\textnormal{V} = \Delta\textnormal{V}_{\textnormal{los}}/\textnormal{sin}(\theta)
\end{equation}
\begin{equation}
    \textnormal{L} = \textnormal{L}_{\textnormal{app}}/\textnormal{cos}(\theta).
\end{equation}

Values of $\theta$ close to 0 imply a very large $\Delta\textnormal{V}$, while values close to 90 imply a very long plume.

For the case of a gravitational disturbance (upper left), we can rule out values of $\theta$ close to 0 because the $\Delta\textnormal{V}_{\textnormal{los}}$ is already at the upper end of the velocities observed for gravitationally disturbed clusters. Larger values of $\theta$ are difficult to rule out in this scenario.

For a jet driven outflow (lower-left), angles close to these two extremes can be ruled out. A long plume would require the jet to act over a very long distance, while a plume with a very large $\Delta\textnormal{V}$ would require an extremely powerful jet.

In the case of a gravitationally disturbed cluster, the true distribution of molecular gas may be more complex than the simple linear arrangement shown in the top left panel of Fig. \ref{fig:A2390_schematic}. Simulations show that the trails resulting from gravitational disturbances are typically bent \citep[e.g.][]{ZuHone2019}. In Fig. \ref{fig:A2390_schematic_curved_plume} we illustrate how the line of sight velocity may show a smooth change, even when the gas has a constant velocity magnitude. However, a long plume of disturbed gas is unable to form unless there is some velocity gradient to induce its elongation. Any plume resulting from a gravitational disturbance is therefore likely to have a velocity structure explained by a mix of the extremes in Figs. \ref{fig:A2390_schematic} and \ref{fig:A2390_schematic_curved_plume}. 

\subsection{Potential Power and Age of the Molecular Plume}
\label{sec:power_and_age}

Given the uncertain origin of the plume, it is of interest to calculate the power required to produce it via an accelerating force such as a jet. Since the velocity field is consistent with linear acceleration from the nucleus outward, the initial velocity, $v_0$, final velocity, $v$, acceleration, $a$, and distance covered, $s$, are related by
\begin{equation}
    a = \frac{v^2 - v_{0}^{2}}{2s},
\end{equation}

which we find to be $6.8\times 10 ^{-14}$ km/s$^2$ (using a velocity change of 250\,km/s and a length of 15\,kpc).

The force driving the velocity change in the plume can be estimated from the product of its mass and acceleration. By using the above acceleration and assuming half of A2390's molecular mass is contained within the plume (i.e. $1.1 \times 10 ^{10}\, \textnormal{M}_{\odot}$, see Table \ref{tab:masses}), we find a value of $1.5 \times 10 ^{35}$ erg/cm.

The kinetic energy of the plume, $3.4 \times 10^{57}$\,erg, is found from the product of the driving force and the distance covered (taking the mean distance covered to be half the plume length). If the plume is made up of gas lifted out from the galaxy centre, considerable work is also done against gravity. Using the isothermal mass profile of Abell 2390 from \citet{Pierre1996}, we find the gravitational potential energy of the plume to be 1.0$\times10^{58}$\,erg. The total kinetic and gravitational potential energy contained within the plume is therefore 1.3$\times10^{58}$\,erg.

The initial velocity, final velocity, acceleration and age of the plume, $t$, are related by
\begin{equation}
    v = v_{0} + at.
\end{equation}

The plume's implied age is therefore 
\begin{equation}
    t = \frac{v - v_{0}}{a},
\end{equation}

\begin{figure*}
	\includegraphics[width=0.9\textwidth]{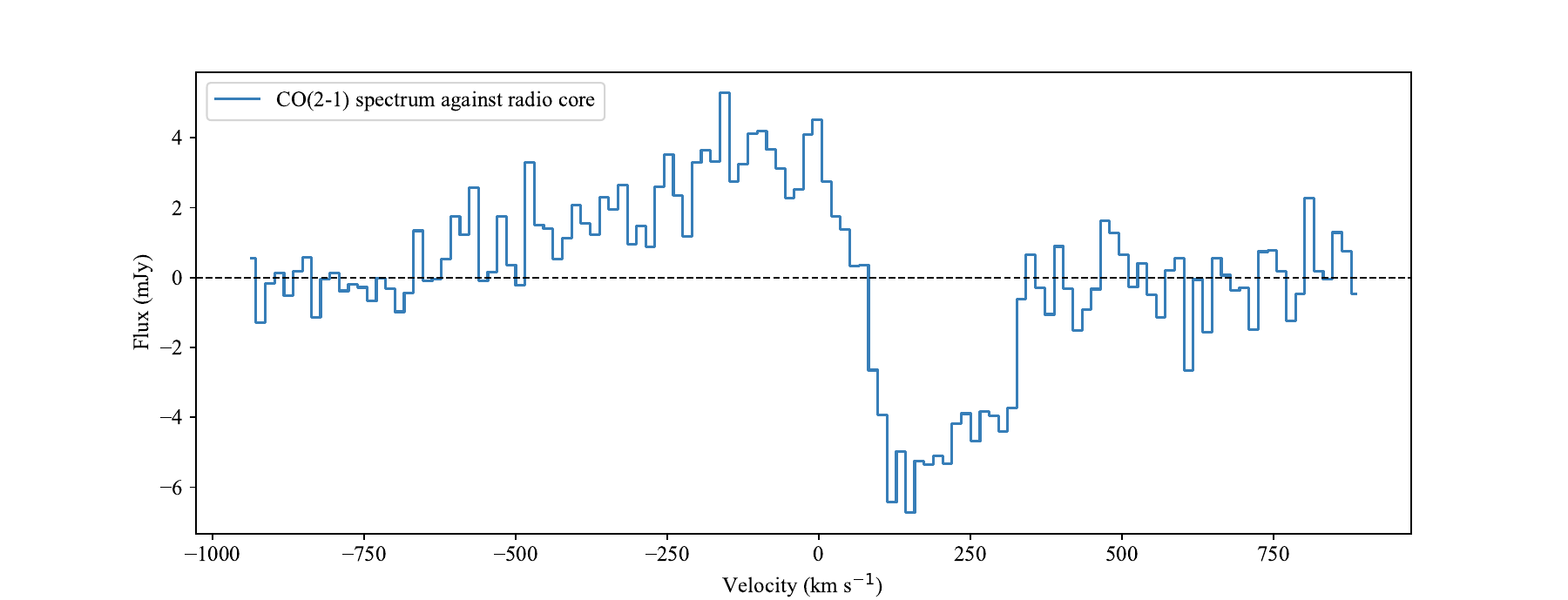}
    \caption{Continuum subtracted CO(2-1) spectrum in Abell 2390. The spectrum is extracted from an approximately beam sized region centred on the radio core. This maximises the visibility of the absorption by excluding much of the spatially extended emission present in Figures \ref{fig:A2390_HST_and_moments_maps} and \ref{fig:pV_diagram}. The zero velocity is set using stellar absorption lines of the core \citep{Hamer2016}. Similarly broad and redshifted CO(1-0) \citep{Rose2019b} and HI absorption \citep{Zahid2004, Hogan_thesis, Hernandez2008} have also been detected against the continuum source.}
    \label{fig:CO(2-1) spectra}
\end{figure*}

\begin{figure}
	\includegraphics[width=\columnwidth]{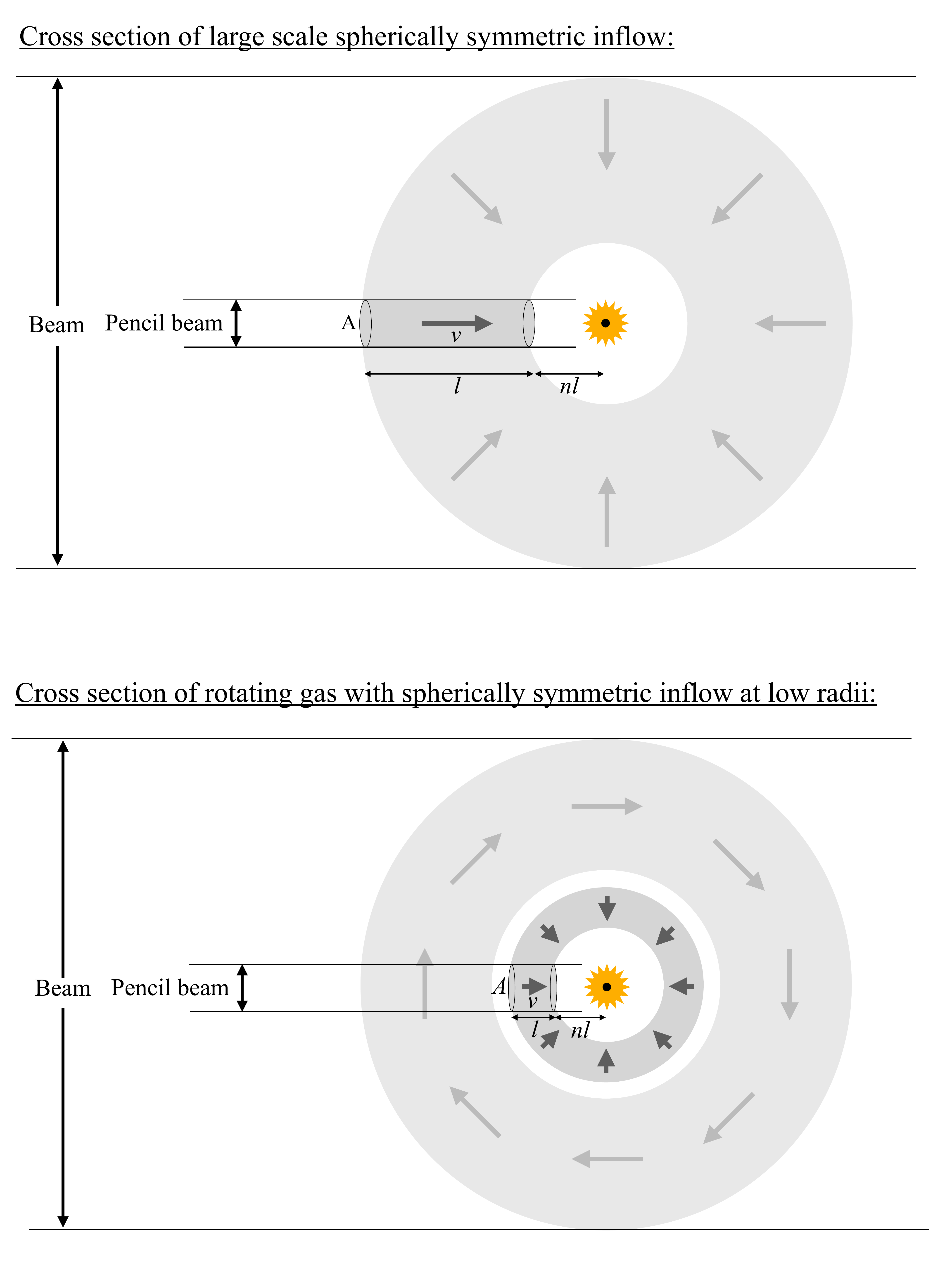}
    \caption{Two simplistic models which could result in the emission-absorption spectrum shown in Fig. \ref{fig:CO(2-1) spectra}. Our line of sight is in the direction from left to right. The top model shows a large scale inflow where the same gas is responsible for both the emission and absorption in the spectrum. The bottom panel shows a model in which the emission is caused by a large scale distribution of gas with some preferred direction of rotation. The absorption is caused by spherically symmetric inflow of molecular gas close to the galaxy centre. Due to the proximity, alignments with the continuum are relatively likely, resulting in the high absorption column density. $l$ is the length of absorbing gas, $v$ is the inflow velocity and $nl$ is the distance from centre to cloud edge expressed in terms of $l$, where $n$ is a constant. $A$ is the cross-sectional area of the pencil beam towards to continuum source. This is much smaller than the observation's beam -- which does not resolve any structure in the continuum.}
    \label{fig:double_schematic}
\end{figure}

\begin{figure}
	\includegraphics[width=\columnwidth]{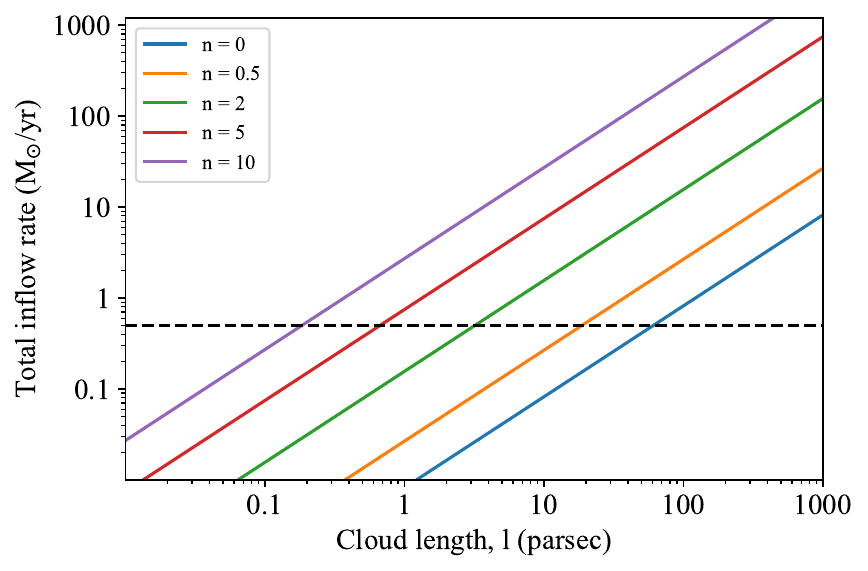}
    \caption{The inferred total inflow rate, $\dot{M}$, as a function of cloud length, $l$, derived for the scenario shown in the lower half of Fig. \ref{fig:double_schematic}. Here, $n$ is a constant, with $nl$ being the distance from the galaxy centre to the near edge of the absorbing gas. The dashed line marks a total inflow rate of 0.5\,M$_{\odot}$/yr.}
    \label{fig:mass_inflow_rates}
\end{figure}

which we find to be $3.7 \times 10^{15}$s, or 0.12\,Gyr. This is of the same order as the cooling time in the galaxy's central 9\,kpc, estimated by \citet{Allen2001} to be 0.4\,Gyr.

By dividing the total energy contained within the plume by its age, we find a power of $3.4 \times 10^{42}$ erg/s. This power is consistent with molecular gas flows commonly found in active central galaxies \citep{Tamhane2022}. In section, \ref{sec:jet_driven_outflow}, we further discuss the possibility of a jet driven outflow.

\citet{Hutchings2000} calculate a star formation rate in Abell 2390's core of $28\, \textnormal{M}_{\odot}$/yr. Values in the \textit{HST} image's two knots (coincident with the plume) are not explicitly stated, but are said to be similar. A star formation rate of $8.8\,\textnormal{M}_{\odot}$/yr has also been estimated by \citet{Rawle2012} from FIR data. At the higher end, a total star formation rate of $80\, \textnormal{M}_{\odot}$/yr implies the galaxy's entire reservoir of molecular gas would be depleted in 0.26\,Gyr. This timescale is similar to the age estimated above, further suggesting that the plume is indeed young and may be sustained by the cooling of hot gas. At the lower end however, a total star formation rate of $8.8\, \textnormal{M}_{\odot}$/yr would deplete Abell 2390's molecular gas in a much longer period of 2.4\,Gyr.

\section{Molecular absorption against radio core}
\label{sec:absorption}

In Fig. \ref{fig:CO(2-1) spectra}, we show a CO(2-1) spectrum extracted from an approximately beam-sized region spatially coincident with Abell 2390's radio core. This contains emission, as well as a 250\,km/s wide absorption feature centred at a velocity of approximately 200\,km/s. 

The absorbing gas lies in front of Abell 2390's radio continuum source, which has some structure at 1--10\,GHz detected by the VLBA \citep{Augusto2006}. However, at 23\,GHz, the continuum is extremely compact, and will be even more so at the frequency of the CO(2-1) line. The absorption can therefore be approximated as coming from a single pencil-beam line of sight. Additionally, the positive sign in the absorption's velocity unambiguously indicates movement of the absorbing gas towards the galaxy centre along our line of sight. It also matches that of previously detected HI and CO(1-0) absorption \citep{Zahid2004, Hogan_thesis, Rose2019b, Hernandez2008}.

As well as being wide, the molecular absorption is deep. The drop in the continuum subtracted flux density is around 6\,mJy/beam, compared with a continuum flux of 8\,mJy/beam. However, in reality the absorption is likely optically thick, with the missing 2\,mJy filled in by CO(2-1) emission. This is confirmed by HCO$^{+}$ observations, which lack molecular emission, but show near saturated absorption (Rose et al. in prep.).

In many respects the absorption against Abell 2390's radio core is consistent with cases where the molecular gas is inferred to be in the central regions of the host galaxy and is moving towards the radio core along the line of sight at hundreds of km/s \citep[e.g. NGC 5044, Abell 2597, S555:][]{David2014, Tremblay2016, Rose2019b, Rose2023}. Specifically, their absorption lines have high inward velocities and are very wide, implying they are from a region with a very high velocity gradient. The absorption lines are also significantly redshifted compared with the molecular emission, suggesting the two are not produced by the same population of clouds. This is in contrast to cases where the absorption is due to chance alignments between the continuum source and the large scale distribution of molecular gas, which results in narrow, low velocity absorption lines contained within the emission \citep[e.g. Hydra-A and IC 4296,][]{Rose2020, Ruffa2019}.

However, Abell 2390 is unique in that it has a deep, wide and highly redshifted absorption feature embedded within the emission. This may be due to the large beam size encapsulating emission from a wide area, but a second possibility is a larger scale inflow. We now discuss this possibility, along with two other explanations for the molecular absorption.

\subsection{Large-scale spherically symmetric inflow}

In the upper panel of Fig. \ref{fig:double_schematic}, we show a simplified model of a large-scale spherically symmetric inflow. In this model, emission is produced by gas on both sides of the galaxy, while absorption against the continuum can only be caused by gas on the near side. This results in a broad emission line, combined with an absorption feature at positive velocities.

In Abell 2390, the line of sight column density of the emission spatially coincident with the radio core is $160\textnormal{M}_{\odot}/\textnormal{pc}^{2}$. For spherically symmetric accretion, the \textit{total} mass inflow rate at a given radius, $\dot{M}$, is:

\begin{equation} \dot{M}=\rho S v \end{equation}

where, $\rho$ is the volume density, $S$ is the surface area of the inflow and $v$ is the inflow velocity.

The total mass within the beam is $0.6\times 10^{10}\, \textnormal{M}_{\odot}$ and the beam radius is 1.9\,kpc, giving a density of $0.2\,\textnormal{M}_{\odot}/\textnormal{pc}^{3}$. This assumes a spherical distribution of gas which extends to the galaxy centre. 

In this model, the blueshifted side of the emission (with negative velocities) will originate from the far side of the galaxy. The intensity weighted mean line of sight inflow velocity for the far sided emission is 250\,km/s. The true inflow velocity is then $$v = \frac{v_{\textnormal{los}}}{ \cos(\theta)},$$ where $\theta$ is the angle made to the line of sight. Evaluating this between $-\frac{\pi}{2}$ and $\frac{\pi}{2}$, the mean inflow velocity is then $v= 390$\,km/s.

The inflow area decreases with the radius, so an approximation is required. If we take $S$ to be the mean area between the centre and the beam's radius, $l$, then $S = \frac{4}{3} \pi l^{2}$. The inferred mass inflow rate is then $1300\textnormal{M}_{\odot}/\textnormal{yr}$ (1sf). This is extremely high and its presence on kiloparsec scales means it would endure for at least several thousand years (since changes cannot take place on a timescale less than the light crossing time). This is orders of magnitude above the Eddington accretion rate, and even when assuming a low accretion-to-jet energy conversion efficiency of $\eta = 1\times10^{-4}$ \citep{Christie2019}, could fuel $7\times10^{45}$\,erg/s jets. This suggests the model of a large scale inflow is not realistic.

\subsection{Spherically symmetric inflow at low radii}

The lower panel of Fig. \ref{fig:double_schematic} shows a second simple model. Here, the emission originates from a massive, kpc scale distribution of molecular gas. Due to its size, this is easily visible in emission. However, assuming it is composed of discrete molecular clouds, its relatively low column density implies that only a very small number of them should align with the continuum. This would produce narrow absorption features \citep[like in Hydra-A][]{Rose2019a}, rather than the single, very wide absorption line we see. In this case therefore, the gas we see in emission cannot also be the cause of the deep and wide absorption. Instead, the absorption may be due to gas close to the galaxy centre which is undergoing spherically symmetric inflow -- this is too low in mass to be visible in emission, but close enough to the continuum source that alignments with our line of sight are reasonably likely. The proximity to the nucleus would also give the molecular gas a large velocity dispersion, explaining the high velocity width of the absorption. 

The emission from gas around the radio core is not resolved in our observations due to the large beam size and there are many potential velocity structures which could produce the broad emission present in Fig. \ref{fig:CO(2-1) spectra}. The exact form is not critical to the model so long as it produces a broad emission line, though one possibility which we represent in Fig. \ref{fig:double_schematic} is a rotating molecular disc like in Hydra-A, where emission is present from -400 to +400\,km/s \citep{Rose2019a}.

Using this model, we can then  perform calculations to estimate the mass accretion rate it implies. With $l$ and $nl$ as defined in Fig. \ref{fig:double_schematic}, the mass flow rate \textit{along the pencil beam line of sight} to the continuum, $\dot{m}$, is:

\begin{equation} 
\dot{m}=\rho A v=\frac{\sigma}{l} A v, 
\label{eq:mass_inflow_of_pencil_beam}
\end{equation}

where $A$ is the cross-sectional area of the continuum, $\rho$ is the volume density, $\sigma$ is the line of sight column density and $v$ is the inflow velocity.

The total spherical mass inflow, $\dot{M}$, and the line of sight inflow, $\dot{m}$, are proportional according to their areas. The surface area of the cloud segment is $A$. The surface area of the sphere (i.e. the total inflow area) changes with radius. However, the mean value is $\frac{4\pi}{3} l^2 (3n^2+3n+1)$. This is found by integrating the function for the area from $nl$ to $nl + l$ and dividing by the range ($l$). With these two functions for the area, $\dot{M}$ and $\dot{m}$ are related by:

\begin{equation} \dot{M}=\dot{m}\frac{4\pi}{3A} (3n^2+3n+1)l^2 \end{equation}

The detection of strong and wide absorption against Abell 2390's radio continuum is unusual, so the absorption along this line of sight is not likely to be representative of an overall accretion rate. Therefore, we can also introduce a factor, $f$, which accounts for `over-detection' along the line of sight. For example, $f=2$ would mean twice as much gas is inflowing along our line of sight as is average. Then: 

\begin{equation} 
\dot{M}=\dot{m}\frac{4\pi}{3A} (3n^2+3n+1)l^2\frac{1}{f}
\label{eq:M_dot}
\end{equation}

Substituting equation \ref{eq:mass_inflow_of_pencil_beam} into \ref{eq:M_dot} gives:

\begin{equation}
\label{eq:mass_inflow_rate}
\begin{split}
\dot{M}&= \frac{\sigma}{l} Av \frac{4\pi}{3A} (3n^2+3n+1)l^2\frac{1}{f}\\
&=\frac{4\pi\sigma v}{3f} (3n^2+3n+1)l
\\
\end{split}
\end{equation}

\begin{figure}
	\includegraphics[width=0.9\columnwidth]{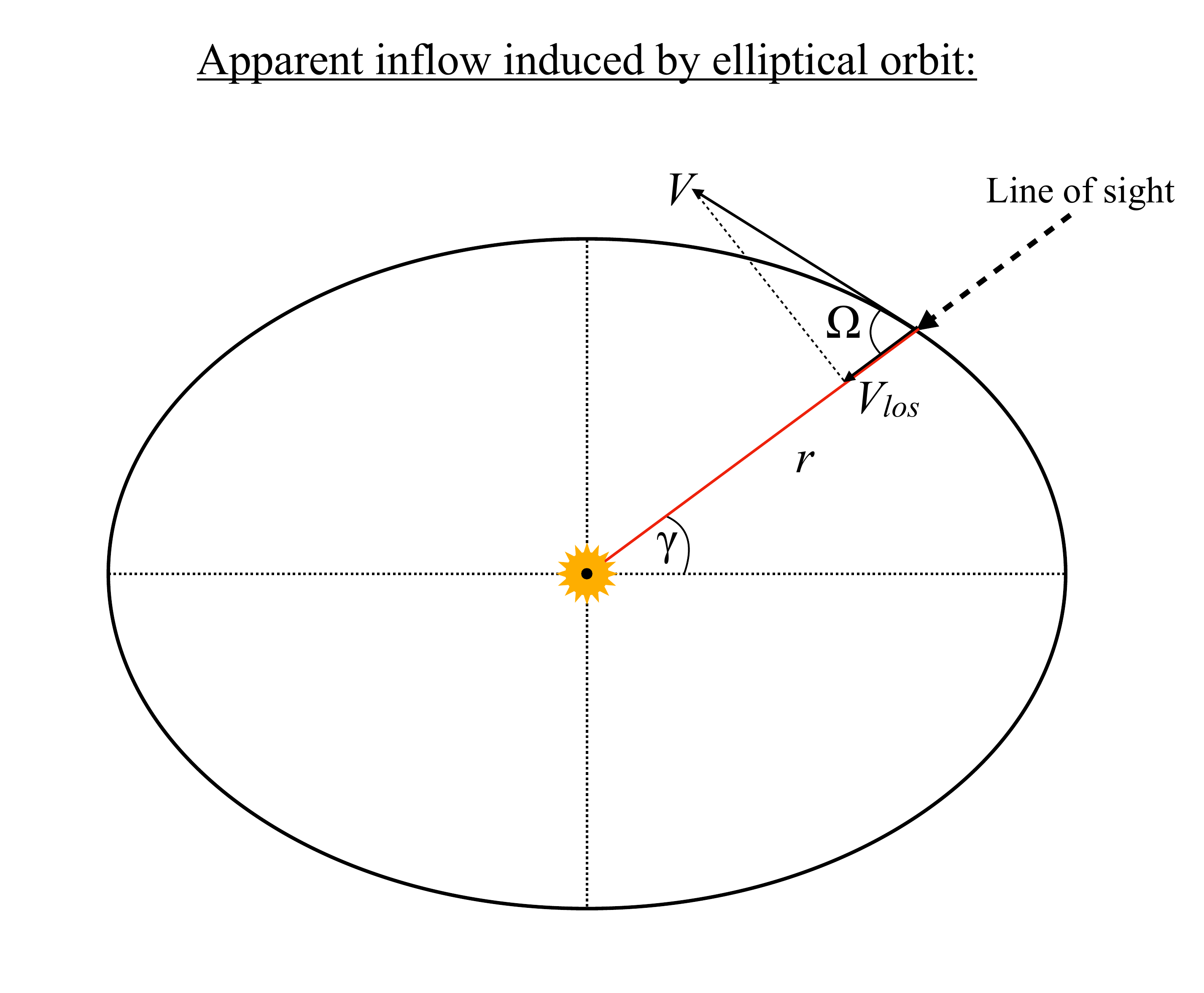}
    \caption{The line of sight velocities of the absorbing gas in Fig. \ref{fig:CO(2-1) spectra} may be a result of elliptical motion around the galaxy centre, with the active galactic nucleus at or close to the centre of the ellipse.}
    \label{fig:elliptical_orbits}
\end{figure} 

We can estimate $\dot{M}$ from our molecular absorption profile. The line of sight column density, $N_{\textnormal{tot}}$, of an optically thin molecular absorption region is:

\begin{equation}
\label{eq:thin_colum_density}
N_{\textnormal{tot}}^{\textnormal{thin}} = Q(T) \frac{8 \pi \nu_{ul}^{3}}{c^{3}}\frac{g_{l}}{g_{u}}\frac{1}{A_{ul}} \frac{1}{ 1 - e^{-h\nu_{ul}/k T_{\textnormal{ex}}}}\int \tau_{ul}~dv~,
\end{equation}

where $Q$($T$) is the partition function, $c$ is the speed of light, $A_{ul}$ is the Einstein coefficient of the observed transition and $g$ the level degeneracy, with the subscripts $u$ and $l$ representing the upper and lower levels \citep{Godard2010,Magnum2015}. The assumption of optically thin absorption in Equation \ref{eq:thin_colum_density} is inappropriate for our CO(2-1) absorption. To account for this, a correction factor can be applied \citep{Magnum2015} to give the following more accurate column density:

\begin{equation}
N_{\textnormal{tot}} = N_{\textnormal{tot}}^{\textnormal{thin}} \frac{\tau}{1-\exp({-\tau})}.
\label{eq:true_column_densities}
\end{equation}

From the above, we find a velocity integrated optical depth of $\tau dv = 214$\,km/s, which gives a column density of $\sigma = 130\,\textnormal{M}_{\odot}/\textnormal{pc}^{2}$ \citep{Rose2023} (assuming a CO/H$_{2}$ ratio of $10^{-4}$ and excitation temperature of 5.5\,$K$).

The absorption in Abell 2390 was initially found as part of a sample of 20 massive galaxies with the highest available X-ray brightness \citep{Rose2019b}. \citet{Rose2023} suggest that 3 detections from those 20 observations are likely due to gas in close proximity to the galaxy centre, so our best estimate is $f = 20/3$. However, Abell 2390 has an absorption column density around twice the value of those similar detections, so the best estimate we can make is that $f = 40/3$ in this case, though there is a high degree of uncertainty. The last remaining constant needed for Equation \ref{eq:mass_inflow_rate} is $v$, shown in Fig. \ref{fig:CO(2-1) spectra} to be $v=200$km/s. With these constants, we find:

\begin{equation}
\dot{M} = 0.0082\,(3n^2+3n+1)l \,\, \textnormal{M}_{\odot}/\textnormal{yr}/\textnormal{pc}
\end{equation}

We can then determine the implied inflow rate for various $n$ and $l$, which we show in Fig. \ref{fig:mass_inflow_rates}. We can also work backwards and assume a reasonable supermassive black hole accretion rate to infer $n$ and $l$. For this reason, in Fig. \ref{fig:mass_inflow_rates} we show a dashed line marking an accretion rate of 0.5\,M$_{\odot}$/yr -- typical of brightest cluster galaxies like Abell 2390. 

The case where $n=0$ is the simplest to consider. This is when the absorbing gas extends to the galaxy centre, and also maximises how far the cloud edge is from the galaxy centre. This suggests the absorbing gas extends to around 70\,pc. For $n=0.5$, the distance between the radio core and nearest edge of the cloud is equal to half the cloud diameter. In this case, $l = 20$\,pc and $nl = 10$\,pc. For $n=10$, $l = 0.15$\,pc and $nl = 1.5$\,pc. If correct, this model therefore suggests Abell 2390's absorbing gas is at most a few 10s of parsecs away from galaxy centre. Under direct inflow and with an average velocity of 200\,km/s, a distance of 10 parsec can be covered in $4\times10^{4}$ years.

\subsection{Apparent inflow induced by elliptical orbits}

As shown in Fig. \ref{fig:elliptical_orbits}, apparent inflow can arise due to elliptical motion. This provides a third possible explanation for Abell 2390's redshifted absorption.

With a mass, $M$, enclosed within an elliptical orbit, the magnitude of the velocity of an object on the orbit is:

\begin{equation}
    V = \sqrt{\textnormal{G}M \left( \frac{2}{r} - \frac{1}{a} \right)},
\end{equation}

where $r$ is the distance between the ellipse centre and the object (as shown in Fig. \ref{fig:elliptical_orbits}) and $a$ is the semi-major axis.

We may define $\gamma$ as the angle between the semi-major axis, the ellipse centre, and the object (see Fig. \ref{fig:elliptical_orbits}). Accordingly, $a$ and $r$ are related by a constant, $k(e,\gamma)$, which depends on the orbit's eccentricity, $e$, and $\gamma$. Therefore, it is convenient to write:

\begin{equation}
    V = \sqrt{\textnormal{G}M \left( \frac{2}{r} - \frac{1}{kr} \right)}\,,
\label{eq:V_ofelliptical_orbit}
\end{equation}

The measured line of sight velocity, $V_{\textnormal{los}}$, and $V$ are related by:

\begin{equation}
    V = \frac{V_{\textnormal{los}}}{\cos{\Omega}},
\label{eq:V_V_los}
\end{equation}

with $\Omega$ as defined in Fig. \ref{fig:elliptical_orbits}.

Combining equations \ref{eq:V_ofelliptical_orbit} and \ref{eq:V_V_los} and solving for $r$, we find:

\begin{equation}
    r = \textnormal{G}M \left( 2 - \frac{1}{k}\right) \left(\frac{\cos{\Omega}}{V_{\textnormal{los}}}\right)^{2}
\end{equation}

This does not account for inclination of the elliptical orbit, which would reduce $V_{\textnormal{los}}$ further. This model will therefore give the maximum possible cloud-supermassive black hole distance for a given $V_{\textnormal{los}}$ and $M$.

Typical eccentricities for the $\sim10$\,pc surrounding Sgr A* are in the range 0.2 - 0.8 \citep{Ferriere2012, Mapelli2016}. For an elliptical orbit of eccentricity of 0.6, viewed at an angle $\gamma = 45^{\circ}$, a line of sight velocity of $v_{\textnormal{los}} = 200$\,km/s (which we see in Abell 2390) can be attained by a range of cloud supermassive black hole distances, $r$ and enclosed masses, $M$. If the central supermassive black hole mass is $10^{9}\textnormal{M}_{\odot}$ and dominates the enclosed mass, then $r = 4.8$\,parsec. If the mass is of order $10^{10}\textnormal{M}_{\odot}$ as work by \citet{Hlavacek-Larrondo2012} implies is likely, we find $r = 48$\,parsec.

Without a $M(r)$ profile at lower radii, it is hard to constrain the location of the absorbing gas in this scenario. Nevertheless, the 200\,km/s line of sight velocity observed in the spectrum of Fig. \ref{fig:CO(2-1) spectra} is clearly attainable in this way. However, the absorption's high velocity dispersion and column density still suggests it is likely close the galaxy centre and the supermassive black hole's sphere of influence, even if there is no inflow.

\section{The origin of Abell 2390's multiphase plume}
\label{sec:discussion}

\subsection{A past gravitational disturbance in Abell 2390?}

Clusters of galaxies frequently experience gravitational disturbances though mergers with small groups and individual galaxies. These can have a significant impact, even when the acquired system is comparatively small \citep[][]{Mazzotta2008, ZuHone2010, ZuHone2013}. The many components of a galaxy and its cluster -- such as the stars, molecular gas, dark matter and the intracluster medium -- are affected differently by gravitational disturbances. Specifically, the dark matter, dense baryonic matter (e.g. stars) and diffuse baryonic matter (e.g. the ICM) can separate from each other due to ram pressure forces \citep[see for example the extreme case of the Bullet Cluster,][]{Markevitch2004}. After their displacement, the diffuse baryonic components orbit the dark matter dominated gravitational potential in an oscillating manner. The characteristic spiral pattern this can create in the X-ray, molecular and H$\alpha$ emission at the centre of a cluster is often referred to as `sloshing'. This activity can mimic uplift by displacing gas from the galactic center along a plume \citep{Hamer2016}.

The velocity profile we see in Abell 2390's molecular gas is similar to several others in which an offset in the X-ray peak and extended optical line emission have been linked to activity of this kind, e.g. Abell 1991, Abell 3444 \citep{Hamer2012} and Abell 2533, which has a 40\,kpc long filament of line emission \citep[][]{Green_thesis}. RX J0821+0752 also has an 8\,kpc long plume of molecular emission and an offset in its X-ray peak \citep{Vantyghem2019}. Other examples include Abell 2029 \citep{Clarke2004}, M87 \citep{Simionescu2010} and Abell 795 \citep{ Ubertosi2021}.

After subtracting elliptical fits to Abell 2390's X-ray emission, tentative cavities and/or sloshing activity have been inferred \citep[][plus our own analysis]{Sonkamble2015,Shin2016}. On larger scales, \citet{Allen2001} identify substructure in the cluster's X-ray emission, suggesting it has not fully relaxed following a recent merger. Similarly, the temperature distribution derived from \textit{XMM} observations is asymmetric, which also favours a recent merger (and therefore ongoing oscillation about the gravitational centre). \citet[][]{Sonkamble2015} also suggest the systematically increasing temperature and metallicity of gas clumps support a prior merger from which the cluster has not fully relaxed \citep[similar to, e.g.][]{Blanton2004}. Finally, in Fig. \ref{fig:X-ray_plume}, we show oscillation in the X-ray brightness as a function of radius. This appears to match that predicted by simulations \citep{ZuHone2010}.

\subsubsection{A trailing plume of molecular gas?}

A past gravitational disturbance may be consistent with the distribution of molecular gas seen in Fig. \ref{fig:A2390_HST_and_moments_maps}, where the brightest, spheroidal component of molecular gas is slightly offset to the north-west of the continuum source -- the same direction as the 15\,kpc long molecular plume. In this scenario, this may be viewed as the molecular gas trailing behind the high density core, which has been efficiently accelerated by the gravitational disturbance due to its minimal drag force. As we show in Fig. \ref{fig:A2390_schematic_curved_plume}, some of the velocity change along the plume may be a simple projection effect. However, a velocity gradient along the plume is still required for it to become elongated.

The gas experiencing the strongest drag forces would end up at the largest distance from the galaxy centre and with the largest velocity offset. The galaxy would also have been in this frame initially, but following its gravitational disturbance, now moves along our line of sight at -250\,km/s (with the negative sign implying movement towards us). This matches what is seen in Fig. \ref{fig:A2390_emission_grid}. The CO(1-0)/CO(2-1) intensity map gives some indication that the the warmer, less dense gas (which would experience the highest drag force) is at the farthest point of the plume and has the largest separation from the galaxy. 

In Fig. \ref{fig:A2390_HST_and_moments_maps}, there is also a region of molecular emission with a velocity dispersion of around 140\,km/s -- at least twice as high as anywhere else. Interestingly, this high dispersion region is of relatively low mass and lies to the opposite side of the continuum source as the plume. This high dispersion region may be akin to a bow wave, behind which calmer molecular gas is trailing.

In this scenario, the small velocity differences between the H$\alpha$ and molecular gas shown in Fig. \ref{fig:Molecular_minus_halpha_velocity} may arise due to the differing ram forces acting on the molecular gas and diffuse hot gas.

\subsubsection{A condensing plume of molecular gas?}

Abell 2390's current star formation rate implies that without further cooling, its molecular gas will be depleted in $0.26-2.4\,\text{Gyr}$ (see section \ref{sec:power_and_age}). At the lower end, this is comparable to the estimated age of the plume. Therefore, the molecular gas may have a short lifespan i.e. the majority of the gas we detect may have formed after the purported gravitational disturbance.

To confound this, we also see evidence of significant ongoing molecular gas formation. \citet{Allen2001} estimate a cooling time in the central $9\,\text{kpc}$ of $4\times10^{8}\,\text{yr}$, giving a cooling rate of $50\,\text{M}_{\odot}/\text{yr}$. In the absence of star formation, this cooling rate would provide a molecular mass equal to that of the plume in 0.2\,Gyr -- compared with its estimated age of 0.12\,Gyr (section \ref{sec:power_and_age}). There is a also a significant amount of hot gas available for cooling. Earlier, we estimated an excess atmospheric gas mass from the \textit{Chandra} X-ray data for the region coincident with the plume of $2\times10^{10}\,\text{M}_{\odot}$. 

The above suggest that much of the molecular gas in Abell 2390's plume may have formed following a gravitational disturbance, rather than having been dragged out by ram pressure forces. The rest frame of this newly cooling gas would be the same as the intracluster medium, but in time would form a gravitationally powered inflow towards the mass centre.

\subsection{A jet driven molecular outflow with lifting by jet inflated bubbles?}
\label{sec:jet_driven_outflow}

Radio observations of Abell 2390 show evidence of multiple epochs of AGN activity. On large scales, the Low-Frequency Array (LOFAR) has shown radio lobes to the east and west with widths of 600 and 300 kpc, respectively \citep[][]{Savini2019}. Much higher angular resolution observations by \citep{Augusto2006} show young radio jets and lobes propagating out in an almost perpendicular direction, suggesting recurring activity from a precessing AGN. The presence of multiple cavities in the X-ray emission at different orientations also points towards AGN feedback having been launched in changing directions. 

From this perspective, it is worth considering if an intermediate period of radio activity may have accelerated Abell 2390's molecular plume. In Sec. \ref{sec:power_and_age}, we estimated that the energy contained within the molecular plume is $1.3\times 10^{58}$\,erg, with a power of $3.4\times 10^{42}$\,erg/s. For brightest cluster galaxies like Abell 2390, this level of power and energy output during a period of AGN activity is often attainable.

However, jet action alone is inconsistent with the velocity gradient, which is highly uniform across the length of the molecular plume (Fig. \ref{fig:pV_diagram}). In H$\alpha$, the plume and its acceleration extend even further, to around 25\,kpc and 600\, km/s. This is an impractically long range for a jet to act over, so the continued acceleration of the plume at these distances would likely require the help of buoyant X-ray cavities. This effect has been observed in Abell 1795, where the southern molecular filament and radio lobe are closely aligned, with gas having been transported out to large radii by buoyant X-ray bubbles \citep{Russell2017}. Highly uniform acceleration in O[II] gas has also been found in Abell 1835, aligned with the galaxy's one-sided 5\,GHz radio emission and trailing behind an X-ray cavity on each side \citep[][Gingras et al. in prep.]{Odea2010}. In Abell 2390, a similar process is a possibility. The plume may have been launched by radio jets and lobes, then further accelerated out to larger radii by rising bubbles. However, we lack the $\sim1$\,GHz observations which would confirm or rule out the existence of these radio lobes. We also see no obvious candidate bubble in the X-ray data, so this would need to have since disintegrated (though bubble remnants may still be visible as the troughs in the X-ray emission of Fig. \ref{fig:X-ray_plume}).

Although a jet driven outflow is plausible for some of the reasons given above, it is not clear why this scenario would produce such an asymmetric distribution of molecular gas. Further, with a jet driven and cavity lifted outflow, the velocity difference between the molecular and H$\alpha$ (Fig. \ref{fig:Molecular_minus_halpha_velocity}) may be easily explained due to a difference in the inertia of the two phases, i.e. one phase being easier to accelerate than the other. However, the molecular phase would typically be the more dense and harder to accelerate, but we find this to be the phase which is receding fastest at the farthest point of the plume.

\subsection{Inflowing molecular gas?}
\label{sec:jet_driven_outflow}

The Abell 2390 cluster is extremely massive, which results in a strong gravitational force towards the cluster centre and its central galaxy. Cooling gas from the intracluster medium can therefore be attracted towards the cluster centre, in what is termed a cooling flow \citep{Fabian1994}. Given that Abell 2390 has a significant cooling rate of $200-300\,\textnormal{M}_{\odot}\textnormal{yr}^{-1}$ \citep[][]{Allen2001}, cooling and inflowing gas is a plausible explanation for the plume. This high cooling rate could provide a molecular mass equal to that of the plume in less than 0.1\,Gyr.

As we show in Fig. \ref{fig:A2390_schematic}, an inflowing plume may lie on either the galaxy's far or near side. In the case of a far sided plume, a line of sight deceleration of the inflowing gas is required. However, as shown in Fig. \ref{fig:A2390_schematic_curved_plume}, this apparent deceleration is achievable without any true reduction in speed if the plume is curved.

\section{Conclusions}

We present new ALMA CO(2-1) observations of molecular gas in the Abell 2390 brightest cluster galaxy. We analyse our ALMA data alongside \textit{HST}, MUSE and \textit{Chandra} observations to constrain the potential origins of its massive multiphase plume. Our main findings and conclusions are as follows:

\begin{itemize}
    \item The total molecular mass in Abell 2390 is $(2.2\pm0.2)\times 10^{10}\, \textnormal{M}_{\odot}$, roughly half of which is contained in a 15\,kpc long molecular plume to the north-west with a velocity change of roughly 250\,km/s.
    \item As well as emission, deep and wide CO(2-1) absorption is observed against Abell 2390's radio core. The absorption is very wide -- although the line of sight to the background continuum is a maximum of a few tens of square parsecs in area, the absorption is around a third as wide as the galaxy's entire molecular emission. It also has high inward velocities, and its column density is two orders of magnitude higher than the emission. These properties are similar to cases where absorbing gas is likely in the central regions of the host galaxy and is moving towards the radio core at hundreds of km/s. We construct two plausible models for how the absorption may arise. In both of these models the gas is close the the galaxy centre, so alignments with the continuum are relatively likely. This proximity to the core explains the high absorption width and column density.
    \item Clusters of galaxies frequently experience disruptive mergers and flybys. These can displace the brightest cluster galaxy's baryonic matter from its dark matter, resulting in characteristic spiral patterns in the cluster's X-ray emission. This `sloshing' may explain the velocity structure seen in Abell 2390's molecular gas. Following a gravitational disturbance, drag forces cause the dark matter and AGN to separate from the intracluster medium, leaving a trail of molecular gas. A merging event would imprint a spatially smooth velocity structure on gas trailing as a result of ram pressure stripping, as well as any gas which cools subsequently. In Abell 2390's molecular gas we see a very smooth velocity gradient, a bright region of molecular gas slightly offset from the nucleus, a more extended plume of lower density molecular gas, and a high velocity dispersion region to the opposite side of the continuum source which may be interpreted as a bow wave. These properties are consistent with the plume having been created following a gravitational disturbance.
    \item Radio observations of Abell 2390 reveal multiple epochs of AGN activity, with LOFAR observations detecting radio lobes to the east and west \citep[][]{Savini2019}, and higher-resolution observations showing radio jets and lobes propagating in a direction almost perpendicular to the LOFAR lobes \citep[][]{Augusto2006}. This suggests recurring activity from a precessing AGN. The presence of multiple cavities in the X-ray emission at different orientations further supports a history of AGN feedback operating in changing directions. During a so far undetected intermediate period of radio activity, the resulting jets may have accelerated the plume of molecular gas. The estimated kinetic and gravitational potential energy contained within the molecular plume are similar to the power output of typical AGN in brightest cluster galaxies. However, the smooth velocity gradient out to radii of at least 15\,kpc is unexpected if a highly energetic and compact jet is the driving force. X-ray cavities may therefore have aided the uplift, but if these ever existed, they are no longer visible.
    \item Abell 2390's molecular plume may also be explained as a molecular inflow. This gas may originate from cooling of the intracluster medium, before being attracted to the cluster center by its strong gravitational pull. In this case, the plume may lie on either the galaxy's far or near side.
    \item The asymmetry of the galaxy's molecular gas distribution is not unexpected following a merger event. However, since jet activity from AGN is counterpoised, a similarly strong jet should exist in the opposite direction to the molecular plume. If this counter-jet does or did exist, it has not acted on any detectable quantities of molecular gas. More sensitive $\sim1$\,GHz radio observations at moderate angular resolution are needed to either confirm or rule out an intermediate phase of radio activity which could have accelerated the plume.
\end{itemize}

\section*{Acknowledgements}

We thank the anonymous referee for their helpful comments which improved this paper and its clarity.

T.R. thanks the Waterloo Centre for Astrophysics and generous funding to B.R.M. from the Canadian Space Agency and the National Science and Engineering Research Council of Canada. A.C.E. acknowledges support from STFC grant ST/P00541/1. HRR acknowledges support from an STFC Ernest Rutherford Fellowship and an Anne McLaren Fellowship. P.S. acknowledges support by the ANR grant LYRICS (ANR-16-CE31-0011).

This paper makes use of the following ALMA data: 2017.1.00629.S, 2021.1.00766.S. ALMA is a partnership of ESO (representing its member states), NSF (USA) and NINS (Japan), together with NRC (Canada), MOST and ASIAA (Taiwan), and KASI (Republic of Korea), in cooperation with the Republic of Chile. The Joint ALMA Observatory is operated by ESO, AUI/NRAO and NAOJ. The National Radio Astronomy Observatory is a facility of the National Science Foundation operated under cooperative agreement by Associated Universities, Inc.

This research made use of \texttt{Astropy} \citep{the_astropy_collaboration_astropy_2013,the_astropy_collaboration_astropy_2018}, \texttt{Matplotlib} \citep{hunter_matplotlib_2007}, \texttt{numpy} \citep{walt_numpy_2011,harris_array_2020}, \texttt{Python} \citep{van_rossum_python_2009}, \texttt{Scipy} \citep{jones_scipy_2011,virtanen_scipy_2020} and \texttt{Aplpy} \citep[][]{aplpy}. We thank their developers for maintaining them and making them freely available.

\section*{Data Availability}

All data products presented in this paper are publicly available for download from each observatory's online data archive.



\bibliographystyle{mnras}



\appendix


\bsp	
\label{lastpage}
\end{document}